\documentclass[twocolumn,amsmath,amssymb,prl,aps,showkeys]{revtex4}
\usepackage{graphicx}
\usepackage{amsmath,amssymb}
\usepackage{mathrsfs}
\usepackage{color}
\usepackage{afterpage}
\usepackage[version=3]{mhchem}
\usepackage{natbib}
\usepackage{soul}
\usepackage[caption=false]{subfig}
\usepackage{array}
\usepackage{multirow}
\usepackage{float}
\usepackage[colorlinks, citecolor=blue, linkcolor=blue, bookmarks=false]{hyperref}

\begin{document}

\title{Coupled spin-valley, Rashba effect and hidden persistent spin polarization in WSi$_2$N$_4$ family}

\author{Sajjan Sheoran\footnote{sajjan@physics.iitd.ac.in}, Deepika Gill, Ankita Phutela, Saswata Bhattacharya\footnote{saswata@physics.iitd.ac.in}} 
\affiliation{Department of Physics, Indian Institute of Technology Delhi, New Delhi 110016, India}

\begin{abstract}
	\noindent 
	The new two-dimensional materials, MoSi$_2$N$_4$ and WSi$_2$N$_4$ are experimentally synthesized successfully and various similar structures are predicted theoretically. Here, we report the electronic properties with a special focus on the band splitting in WA$_2$Z$_4$ (A=Si, Ge; Z=N, P, As), using state-of-the-art density functional theory  and many-body perturbation theory (within the framework of G$_0$W$_0$ and BSE). Due to the broken inversion symmetry and strong spin-orbit coupling effects, we detect coupled spin-valley effects at the corners of first Brillouin zone (BZ). Additionally, we observe cubically and linearly split bands around the $\Gamma$ and M points, respectively. Interestingly, the in-plane mirror symmetry ($\sigma_h$) and the reduced symmetry of arbitrary \textbf{\textit{k}}-point, enforce the persistent spin textures (PST) to occur in full BZ. We induce the Rashba splitting by breaking the $\sigma_h$ through an out-of-plane external electric field (EEF). The inversion asymmetric site point group of W atom, introduces the hidden spin polarization in centrosymmetric layered bulk counterparts. Therefore, the spin-layer locking effect, namely, energy degenerate opposite spins spatially segregated in the top and bottom W layers, has been identified. Our low energy \textit{\textbf{k.p}} model demonstrates that the PST along the M-K line is robust to EEF and layer thickness, making them suitable for applications in spintronics and valleytronics.
\end{abstract}
\maketitle

\section{Introduction}
The discovery of graphene has enormously revolutionized the field of atomically thin two-dimensional (2D) materials because of its extraordinary electronic properties compared to the bulk counterparts~\cite{geim2009graphene,novoselov2012roadmap, kane2005quantum}. Recent advancements in experimental techniques have explored many new 2D materials beyond graphene, including germanene~\cite{ni2012tunable}, silicene~\cite{molle2018silicene}, 2D boron nitride~\cite{chopra1995boron,golberg2007boron}, 2D transition metal dichalcogenides (TMDs)~\cite{manzeli20172d,xu2014spin,wang2018colloquium} and transition metal carbides/nitrides (MXenes)~\cite{shuck2020scalable,zhan2020mxene}. The septuple atomic layer 2D materials MoSi$_2$N$_4$ and WSi$_2$N$_4$ have been experimentally synthesized by passivating the surfaces of non-layered materials~\cite{hong2020chemical}. These materials have a semiconducting nature with excellent ambient stability, high strength and considerable carrier mobility. After that, theoretical studies based on density functional theory (DFT) have predicted several similar thermodynamically stable compounds having formula MA$_2$Z$_4$ (M is an early transition metal viz. Cr, Mo, W, V, Nb, Ta, Ti, Zr or Hf; A = Si or Ge; Z = N, P or As)~\cite{wang2021intercalated}. Subsequent studies have revealed that these materials have promising electronic, optical, mechanical, thermal and non-trivial topological properties\cite{yang2021valley,huang2022mosi2n4,mortazavi2021exceptional}. Moreover, straining, twisting and stacking of these 2D layers to form heterostructures and moire superlattices tune their electronic properties~\cite{pacchioni2020valleytronics, kang2013electronic,singh2018structural,he2014stacking}.

The progress in 2D materials also generates the impetus for realizing physical properties desirable in spintronics and valleytronics. In condensed matter physics, discussion of the symmetry in crystalline solids plays a pivotal role in understanding the physical properties. For instance, inversion and time-reversal symmetry together force the bands to be doubly degenerate throughout the Brillouin Zone (BZ). The coupling of inversion symmetry breaking with spin-orbit coupling (SOC) lifts the band degeneracy mainly through Rashba and Dresselhaus effects. However, if time-reversal symmetry breaks along with inversion symmetry breaking, the spin-valley effects are induced. The presence of well-separated multiple energy extremal points in momentum-space, generally referred to as valleys, constitutes a binary index for low energy carriers~\cite{xiao2012coupled}. Valley index is robust to the scattering by smooth deformations and long-wavelength phonons. Therefore, it can be exploited for information processing and encoding. The lifting of band degeneracy results in non-trivial momentum-dependent spin textures, which are mainly governed by the type of splitting. In general, the SOC introducing these spin splitting effects, also introduces two kinds of spin dephasing mechanisms, viz. Dyakonov-Perel~\cite{schmidt2016quantum}  and Elliot-Yafet~\cite{ochoa2012elliot} mechanisms. One of the possible ways to overcome these spin dephasing mechanisms is persistent spin texture (PST), in which spin directions become independent of the crystal momentum (\textbf{\textit{k}})~\cite{schliemann2017colloquium}. PST is obtained by trading off the strength of the linear Rashba and Dresselhaus effects in heterostructures and surfaces. It has also been shown that non-symmorphic space group symmetry could be exploited to obtain PST around specific high symmetry points and paths~\cite{tao2018persistent}. Also, cubic spin splittings can lead to symmetry-protected PST around the zone center in bulk materials~\cite{zhao2020purely}. Analogous to charge, utilizing the spin and valley degrees of freedom through the non-dissipative spin transport is a vital ingredient of spintronics and valleytronics. 

The 2D MoSi$_2$N$_4$, WSi$_2$N$_4$ and MoSi$_2$As$_4$, host massive Dirac fermions with strong spin valley coupling near the corners of BZ. Additionally, the valley fermions manifest valley-contrasting Berry curvature, and valley-selective optical circular dichroism~\cite{li2020valley,yang2021valley,ai2021theoretical}. This can be attributed to the intrinsic broken inversion symmetry coupled with large SOC arising from $d$-orbitals of transition metal. Under an in-plane electric field, carriers in different valleys drift in opposite directions, leading to the valley Hall effect.  MoSi$_2$N$_4$ shows exceptional piezoelectricity, photocatalytic water splitting and high carrier mobility of 270-1200 cm$^2$/Vs, which is better than widely used TMDs~\cite{mortazavi2021exceptional}. Stacking of MoSi$_2$N$_4$ monolayers generates dynamically stable bilayer and bulk materials with thickness-dependent properties~\cite{islam2021tunable}. Most of the studies focus on the valley splitting in monolayers around the corners of the BZ. However, the complete analysis of SOC induced spin splitting is still unclear.

This work reports fully relativistic calculations within the framework of DFT on the SOC-driven spin splitting in WA$_2$Z$_4$ (A=Si, Ge; Z=N, P, As) family. Our calculations are supplemented by the effective \textit{\textbf{k.p}} model derived from the method of invariants. The calculations reveal the PST in the whole region of BZ in a single layer of these materials. The full zone persistent spin texture (FZPST) is protected by the surface inversion symmetry (SIS) \{(x,y,z)$\rightarrow$(x,y,-z)\}. In addition to spin-valley locked splitting around the corners of BZ, we observe a cubic splitting around the center of BZ. Significant splitting around the M point is linear in \textbf{\textit{k}}, which is found to be complementary to the linear Rashba and Dresselhaus effects. It is a well-known fact that SIS hinders the conventional linear Rashba spin splitting which is later observed under the application of out-of-plane electric field. The \textit{\textbf{k.p}} analysis shows that PST along M-K path is robust to the external electric field. It is generally expected that the presence of inversion symmetry in the bulk WSi$_2$N$_4$, causes the spin splitting and polarization to vanish. However, our calculations show the 100\% spin polarization of doubly degenerate bands in centrosymmetric bulk and even layered WSi$_2$N$_4$, which are spatially segregated. We demonstrate that such hidden spin polarization arises by the inversion asymmetric site point group of W atom. Thus, WA$_2$Z$_4$ family broaden the range of currently useful materials in the fields of spintronics and valleytronics.

 \section{Computational Methods}
The geometry optimization and electronic structures are obtained using first-principles  plane-wave calculations via. DFT as implemented in the Vienna $ab$ $initio$ simulation package (VASP)~\cite{kresse1996efficient}. The projector augmented wave (PAW)~\cite{blochl1994projector, kresse1999ultrasoft} method and plane waves with an energy cut-off of 600 eV are incorporated. The generalized gradient algorithm includes exchange and correlation effects using Perdew-Burke-Ernzerhof (PBE)~\cite{perdew1996generalized} functional. The electronic band structure results are further checked by using the hybrid Heyd-Scuseria-Ernzerhof (HSE06)~\cite{perdew1998perdew} functional and G$_0$W$_0$@PBE~\cite{hedin1965new,hybertsen1985first} method. The SOC is considered in all the calculations self-consistently. BZ is sampled using Monkhorst-Pack~\cite{monkhorst1976special} mesh of 15$\times$15$\times$1 and 21$\times$21$\times$1 for geometry optimization and self-consistent calculations, respectively. The convergence criterion for energy is set to 10$^{-6}$ eV. In order to remove spurious interaction between periodically repeated 2D layers, a vacuum slab of 20 {\AA} is inserted along the $z$-axis. At the same time, to consider the vdW interaction between two layers, the Tkatchenko-Scheffler~\cite{tkatchenko2012accurate} method with iterative Hirshfeld partitioning is used. The structures are fully relaxed using the conjugate gradient method till the maximum force on every atom is smaller than 1 meV{\AA}. The dynamical stabilities are calculated using density functional perturbation theory (DFPT) using PHONOPY~\cite{togo2015first} code with a supercell of 3$\times$3$\times$1. The dipole layer method introduces an electric field, and dipole moments are calculated using the Berry phase method~\cite{PhysRevB.47.1651}. We determine the optical properties by considering the many-body perturbation theory (MBPT). Dielectric functions are computed by solving the Bethe-Salpeter equation (BSE)~\cite{albrecht1998ab} on top of G$_0$W$_0$@(PBE+SOC).
\begin{figure*}
	\includegraphics[width=16cm]{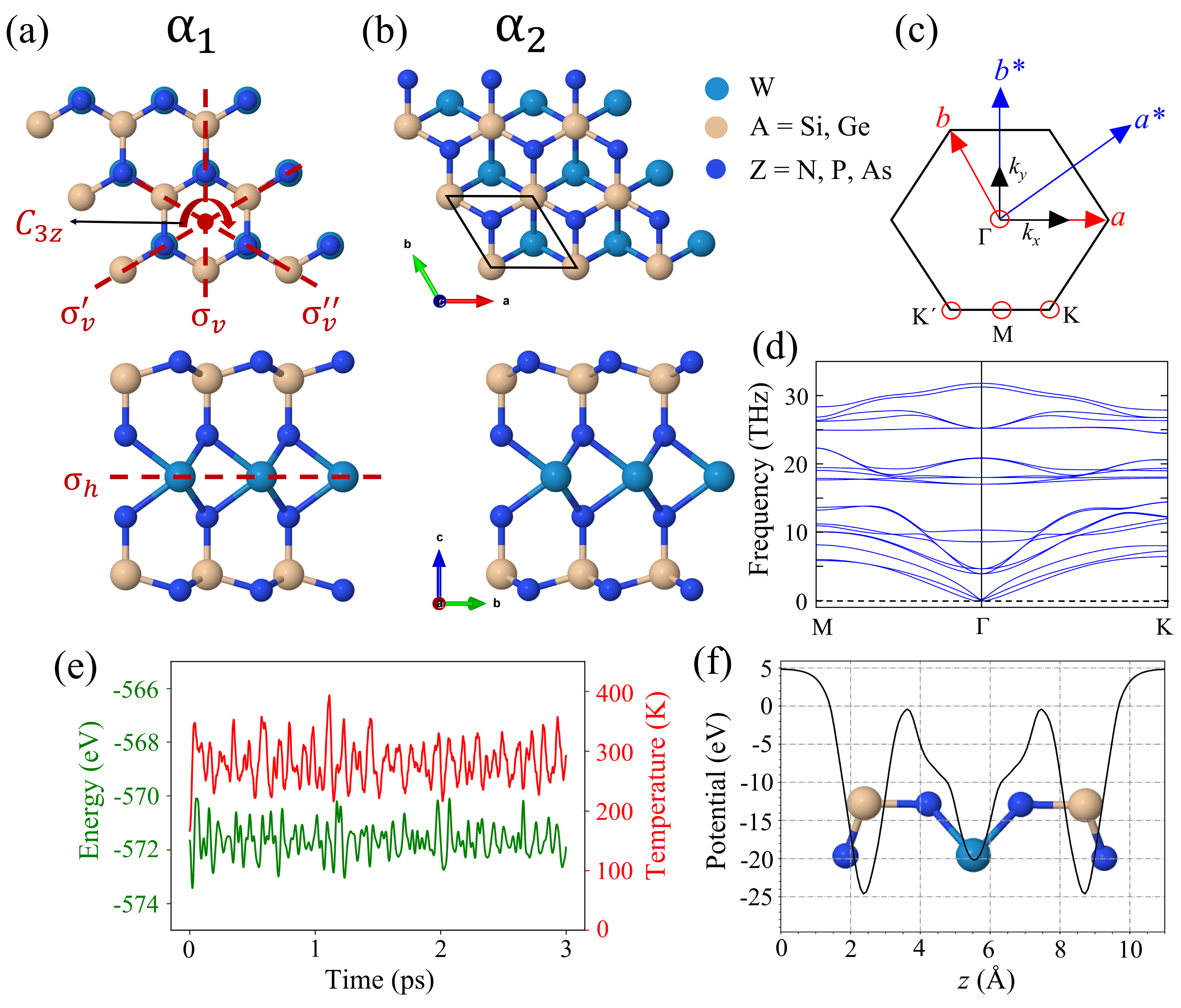}
	\caption{Crystal structures of (a) $\alpha_1$- and (b) $\alpha_2$-WA$_2$Z$_4$ monolayers with symmetry operations highlighted using red dashed line. The solid black lines indicate the corresponding unit cell. (c) The first BZ and the high symmetry \textbf{\textit{k}}-points ($\Gamma$, M, K and K') are indicated using red circles. (d) The phonon band dispersions of $\alpha_2$-WSi$_2$N$_4$ monolayer. (e) Variation of total energy and temperature for $\alpha_2$-WSi$_2$N$_4$ monolayer during AIMD simulation for 3 ps at 300 K. (f) Planar average of the electrostatic potential energy of the $\alpha_2$-WSi$_2$N$_4$ monolayer as a function of out-of-plane axis.}
	\label{p1}
\end{figure*}

\section{Results and Discussion}
\subsection{Structural properties, stability and polarization}
The top and side views of the optimized crystal structure of  WA$_2$Z$_4$ monolayers are shown in Fig.~\ref{p1}(a) and \ref{p1}(b), which can be viewed as the WZ$_2$ layer sandwiched between two A-Z bilayers. Depending on the stacking of A-Z bilayers, different types of crystal structures are possible. Many layered materials are influenced by these stacking effects~\cite{he2014stacking}. In our work, we have considered the two most stable phases i.e., $\alpha_1$ and $\alpha_2$, which differ by the position of N atoms. These two phases share the same hexagonal lattice structure with space group $P\overline{6}m2$. The associated point group symmetry is $D_{3h}$, containing identity operation ($E$), three-fold rotation ($C_{3z}$) having $z$ as the principal axis, mirror operations perpendicular to principal axis ($\sigma_h$) and parallel to principal axis ($\sigma_v$) (see Fig.~\ref{p1}(a)). The optimized lattice constants of WA$_2$Z$_4$ monolayers are provided in Table~\ref{t1}. As expected, the lattice constants increase with increasing atomic number of the group IV and V atoms. Thus, WSi$_2$N$_4$ and WGe$_2$As$_4$ exhibit the smallest and largest lattice constants i.e., 2.88 {\AA} and 3.63 {\AA}, respectively. The same coordination environment in $\alpha_1$ and $\alpha_2$ structures leads to slightly different lattice constants (within 0.5\%). Our lattice constants are in well agreement with already reported experimental and theoretical studies~\cite{hong2020chemical,wang2021intercalated, yang2021valley, huang2022mosi2n4,islam2021tunable}.
\begin{figure*}
	\includegraphics[width=16cm]{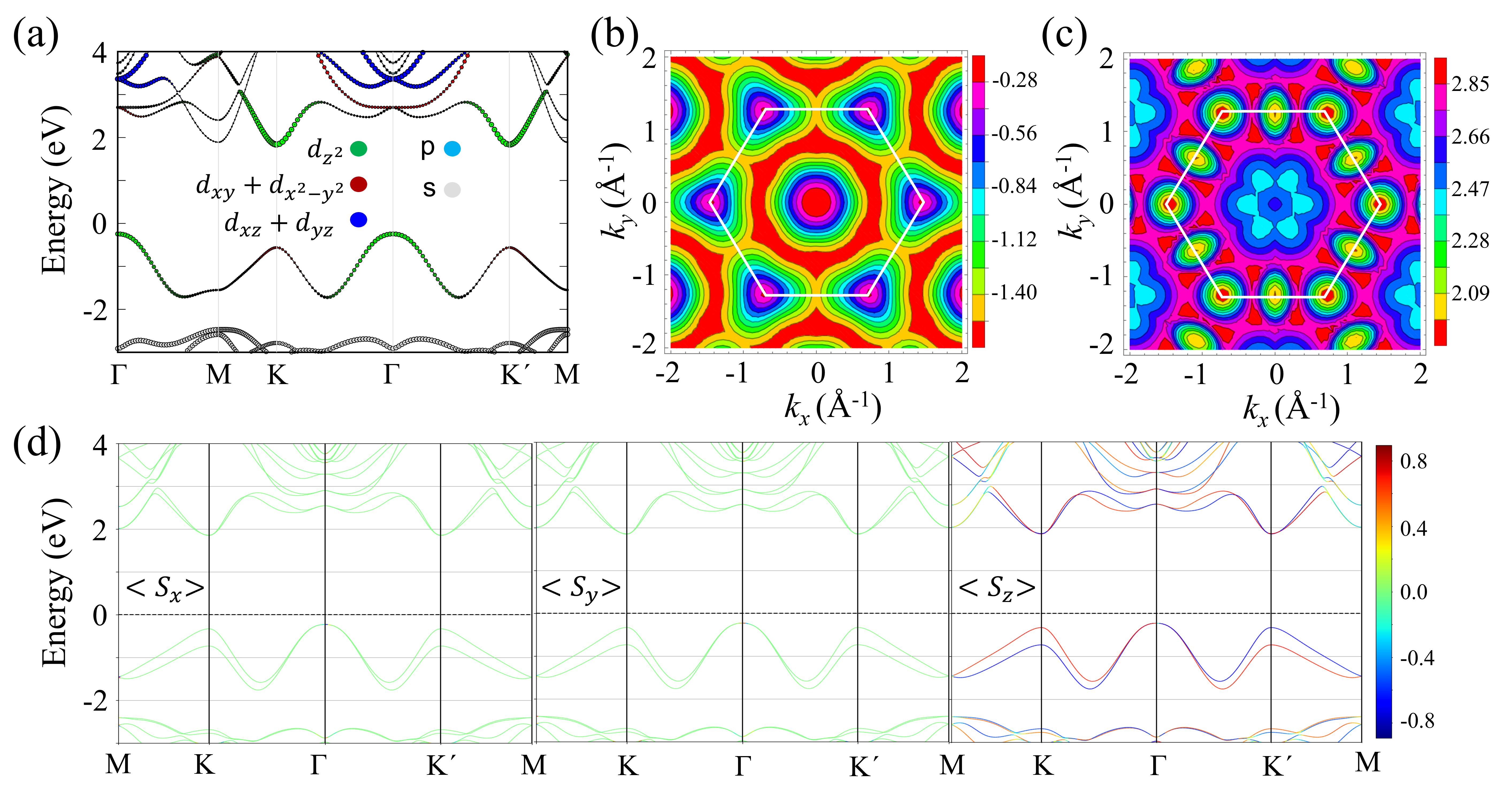}
	\caption{ (a) Orbital-resolved electronic band structure of $\alpha_2$-WSi$_2$N$_4$ monolayer without including SOC. Symbol size is proportional to the particular orbital weightage. The constant energy contours of (b) topmost valence band and (c) lowest conduction band in the momentum-space. The white hexagon is the first BZ. (d) Spin-resolved electronic band structure of $\alpha_2$-WSi$_2$N$_4$ monolayer. The color bar shows the $x$-, $y$- and $z$- components of spin polarization.}
	\label{p2}
\end{figure*}
\begin{table*}
	\begin{center}
		\caption{ Optimized lattice constant ($a=b$) of the unitcell of WA$_2$Z$_4$ monolayers. The calculated formation energy per atom ($E_{for}$), dipole moment per unit cell ($P_y$), band gaps without and with SOC using PBE ($E_g^{\textrm{PBE}}$, $E_g^{\textrm{PBE+SOC}}$) and HSE06 ($E_g^{\textrm{HSE+SOC}}$, $E_g^{\textrm{HSE+SOC}}$). The calculated band gaps using $\textrm{G}_0\textrm{W}_0$ performed on top of PBE functional with inclusion of SOC ($E_g^{\textrm{GW@PBE+SOC}}$).}
		\label{t1}
		\begin{tabular}{p{2.0cm} p{1.4cm} p{1.5cm} p{1.5cm} p{1.5cm} p{1.5cm} p{1.5cm} p{1.5cm}p{1.5cm}p{2.3cm}}
			\hline
			\hline
			& Phase  & $a$  & $E_{for}$ & $P_y$ & $E_g^{\textrm{PBE}}$  & $E_g^{\textrm{PBE+SOC}}$ &$E_g^{\textrm{HSE}}$  &$E_g^{\textrm{HSE+SOC}}$  & $E_g^{\textrm{GW@PBE+SOC}}$  \\
			& &({\AA}) & (eV) & (e{\AA}) & (eV) &  (eV) & (eV) & (eV) & (eV) \\\hline  
			WSi$_2$N$_4$ & $\alpha_1$ &  2.881 & -0.746 & 1.77 & 2.33 & 2.32 & 3.11 & 3.09 & 3.78 \\
			& $\alpha_2$ &  2.894 &  -0.772 & 1.80 & 2.08 & 2.06 & 2.69 & 2.64 & 3.36 \\ 
			WGe$_2$N$_4$ & $\alpha_1$ &  2.976 & -0.532 & 1.82 & 1.68 & 1.64 & 2.13 & 2.05 & 2.76  \\
			& $\alpha_2$ &  2.988 & -0.543 & 1.83 & 1.33 & 1.31 & 1.76 & 1.71 & 2.38 \\ 
			WSi$_2$P$_4$ & $\alpha_1$ &  3.413 & -0.265 & 2.03 & 0.94 & 0.69 & 1.22 & 0.85 & 1.37 \\
			& $\alpha_2$ &  3.423 & -0.252 & 2.04 & 0.55 & 0.32 & 0.87 & 0.76 & 0.93  \\ 
			WGe$_2$P$_4$ & $\alpha_1$ &  3.486 &  -0.292 & 2.31 & 0.70 & 0.45 & 0.94 & 0.56 & 1.02 \\
			& $\alpha_2$ &  3.498 & -0.273 & 2.32 & 0.47 & 0.23 & 0.74 & 0.38 & 0.73 \\ 
			WSi$_2$As$_4$& $\alpha_1$ &  3.561 & -0.079 & 2.11 & 0.78 & 0.52 & 1.03 & 0.60 & 1.02 \\
			& $\alpha_2$ &  3.575 & -0.060& 2.13& 0.50 & 0.22 & 0.77 & 0.39 & 0.67 \\ 
			WGe$_2$As$_4$& $\alpha_1$ &  3.624 & -0.019 & 2.39& 0.60 & 0.33 & 0.85 & 0.38 & 0.79  \\
			& $\alpha_2$ &  3.632 & -0.017 & 2.40 & 0.45 & 0.16 & 0.70 & 0.37 & 0.63 \\ 
			\hline\hline
		\end{tabular}
	\end{center}
\end{table*}

To confirm the energy feasibility of WA$_2$Z$_4$ monolayers, the formation energy per atom ($E_{for}$) is calculated as follows
\begin{equation}
	E_{for}=\{E_{tot}-(n_WE_W+n_AE_A+n_ZE_Z)\}/(n_W+n_A+n_Z)
\end{equation}
where $E_{tot}$ is the total ground state energy of WA$_2$Z$_4$ monolayer. $E_W$, $E_A$ and $E_Z$ are the chemical potentials of isolated W, A and Z atoms, respectively. $n_W$, $n_A$ and $n_Z$ are the total number of W, A and Z atoms in the unit cell, respectively. As seen from Table~\ref{t1}, the negative value of $E_{for}$ shows the energy feasibility of WA$_2$Z$_4$ monolayers. The WA$_2$N$_4$ is most stable in $\alpha_2$ phase, while $\alpha_1$ phase has the lowest energy in the case of WA$_2$P$_4$ and WA$_2$As$_4$, which is consistent with the previous reports~\cite{wang2021intercalated}. However, some previous studies concern only the  $\alpha_2$-phase for WA$_2$P$_4$ and WA$_2$As$_4$~\cite{yang2021valley, huang2022mosi2n4,islam2021tunable}. Therefore, the complete analysis of both phases is more realistic approach. Further, to examine the dynamical stability, we have plotted the phonon dispersion plots for the concerned compounds (see Fig.~\ref{p1}(d)). The unit cell of WA$_2$Z$_4$ monolayer consists of seven atoms (see Fig.~\ref{p1}(b)). As a consequence, phonon plots show three acoustic and eighteen optical branches. As the general property of 2D materials, quadratic dispersion is found in out-of-plane ZA mode around the $\Gamma$ point. The absence of imaginary frequencies in phonon plots confirms the dynamical stability of the WA$_2$Z$_4$ monolayers. To further check the room temperature stability, $ab$ $initio$ molecular dynamics (AIMD) simulations are performed at 300 K for 3 ps with a time step of 1 fs (see Fig.~\ref{p1}(e)). The resultant small energy, temperature fluctuations and no structural disruption, confirm the thermal stability of WA$_2$Z$_4$ monolayers.

Polarization properties play a critical role in determining the intrinsic electronic properties, therefore, we investigate these effects in WA$_2$Z$_4$. Firstly, charge transfer using Bader techniques is performed. Significant charge transfer is found in these monolayers, where Z atoms receive charge from W and A atoms. Different electronegativities lead to different amount of charge received by different Z atoms. For instance, outer N gets a charge of 2.23 e$^-$ from Si, whereas As receives only 0.75 e$^-$. Due to SIS, charge transfer is same for the innermost and outermost layers. Therefore, planar average electrostatic potential shown in Fig.~\ref{p1}(f) is symmetric along $z$-direction, centering W atom. Furthermore, the only in-plane electric dipole moment (along $y$-direction) is observed (see Table~\ref{t1}).

\subsection{Electronic structure and spin splitting} 
As the physics in WA$_2$Z$_4$ monolayers is essentially the same, therefore, taking $\alpha_2$-WSi$_2$N$_4$ as a representative, we investigate the electronic properties of this class of materials. Figure~\ref{p2}(a) shows the orbital-projected electronic band structure of $\alpha_2$-WSi$_2$N$_4$ without SOC calculated using GGA-PBE functional. It has an indirect band gap of 2.08 eV, which is consistent with the previous findings~\cite{hong2020chemical,wang2021intercalated, huang2022mosi2n4,islam2021tunable}. The W-$d$ orbitals dominate the states near band edges. The valence band maximum (VBM) at $\Gamma$ and conduction band minimum (CBm) at K point mainly consist of $d_{z^2}$ orbitals, whereas, $d_{xy}$ and $d_{x^2-y^2}$  contribute to the top valence states at K/K$^\prime$ point. To clearly visualize these two bands, constant energy contours are plotted in momentum-space, as shown in Fig.~\ref{p2}(b) and \ref{p2}(c). It is found that the energies near CBm and VBM are elliptically wrapped. On the other hand, energies are triangularly wrapped for the valence band around K/K$^\prime$ point, similar to the MoS$_2$ family~\cite{xiao2012coupled}. In addition to WSi$_2$N$_4$, WGe$_2$N$_4$ also has an indirect band gap with band edge positions similar to WSi$_2$N$_4$. On contrary, WA$_2$P$_4$ and WA$_2$As$_4$ have a direct band gap at K/K$^\prime$ point (see sections I and II of Supplemental Material (SM)). 

When SOC is included, as seen in Fig.~\ref{p2}(d), the band structure is strongly modified and a sizable spin splitting is observed throughout the BZ, which is the consequence of intrinsic inversion symmetry breaking. The band gap of $\alpha_2$-WSi$_2$N$_4$ decreases to 2.06 eV owing to the SOC induced spin splitting. As PBE is known to underestimate the electronic band gap, the band structures are also calculated using more sophisticated functional HSE06 and G$_0$W$_0$ approach (see sections I and II of SM). The values of band gaps using different functionals are compared in Table~\ref{t1}. To appraise the optical absorption of WSi$_2$N$_4$, we calculate the real and imaginary parts of the dielectric tensor. The dielectric tensor ($\epsilon$) is isotropic in plane due to the $D_{3h}$ point group symmetry (see section II of SM). Therefore, we consider only the $\epsilon_{xx}$ by performing BSE calculation on top of G$_0$W$_0$@PBE+SOC as shown in Fig. S4. The first absorption peak is observed at 2.49 eV, therefore, it is responsive to the energy spectrum in the visible region. 
\begin{figure}
	\includegraphics[width=8.5cm]{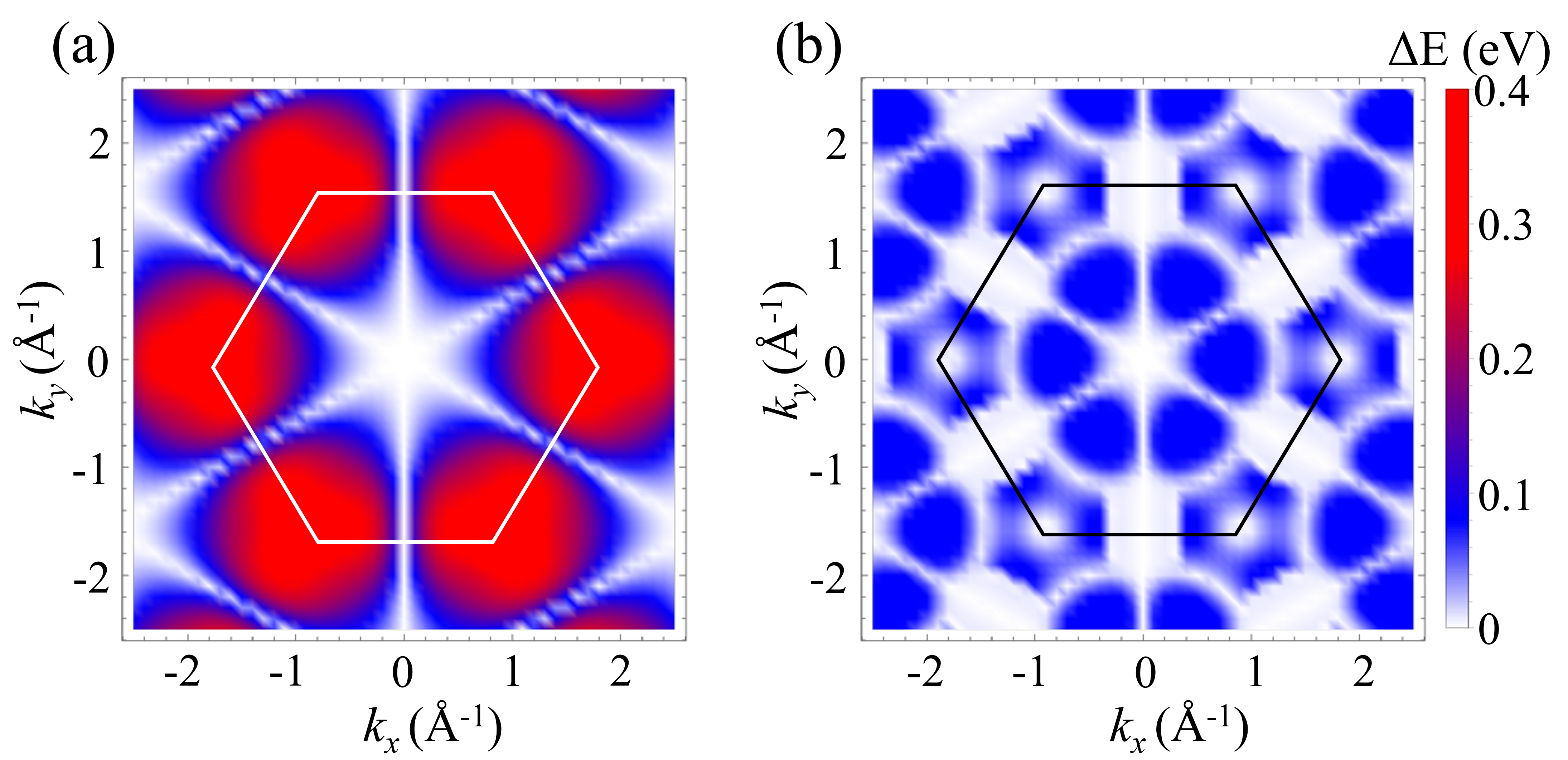}
	\caption{ The spin splitting energy ($\Delta E=|E(k,\uparrow)-E(k,\downarrow)|$) for (a) topmost valence band and (b) lowest conduction band mapped over the full BZ.}
	\label{p3}
\end{figure}

The transport properties of electrons are mainly governed by the conduction and valence bands. Figure~\ref{p3} shows the magnitude of spin splitting energy ($\Delta E=|E(k,\uparrow)-E(k,\downarrow)|$) for the topmost valence bands and lowest conduction bands in the entire momentum-space. Splitting present in the valence band is significantly larger as compared to the conduction band. For this reason, we mainly focus on the valence bands. The splitting is observed in the full BZ except for the high symmetry path $\Gamma-$M, where the eigenstates are at least two-fold degenerate. The degenerate bands split around the time-reversal invariant $\Gamma$ and M points. Importantly, splitting observed around M and $\Gamma$ point is anisotropic in nature. The maximum spin splitting energy observed is up to the order of 35 eV along the M$-$K line, which  mainly originates from the strong hybridization of $d_{xy}$ and $d_{x^2-y^2}$ orbitals. Similar features of spin splitting are also observed in the whole WA$_2$Z$_4$ family (see Fig. S1 in SM). The DFT results are further examined using symmetry analysis.
\begin{figure*}
	\includegraphics[width=16cm]{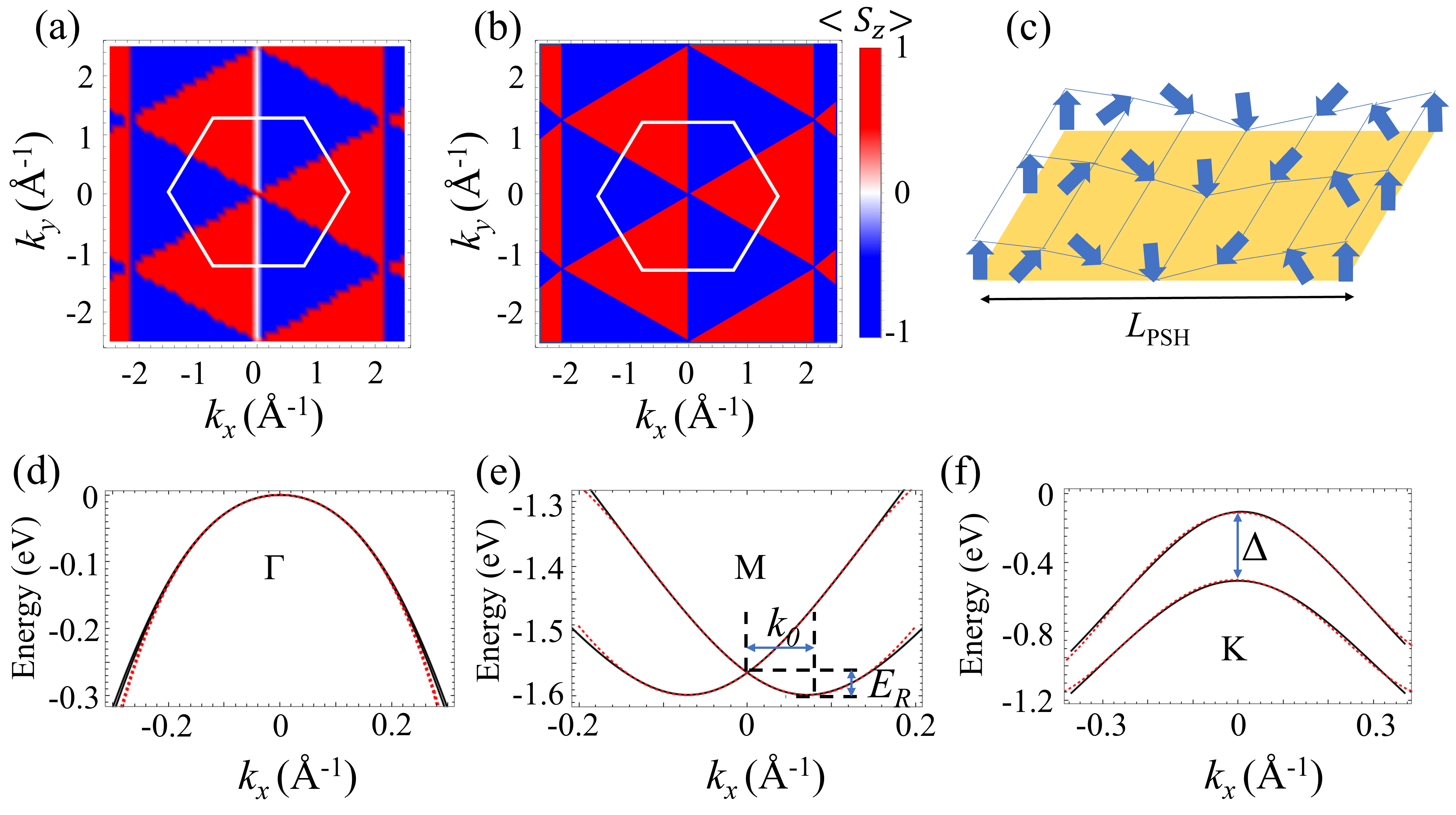}
	\caption{ The spin textures computed using (a) DFT and (b) \textit{\textbf{k.p}} models. The color bar shows the out-of-plane ($z$-) component  of spin textures. (c) Schematic diagram of PSH state. The $L_{\textrm{PSH}}$ represents the wavelength of spatially periodic mode. The band structure of topmost valence bands around (d) $\Gamma$, (e) M and (f) K points. Here, the black solid lines and red dots represent the band structure computed using DFT and \textit{\textbf{k.p}} models, respectively.}
	\label{p4}
\end{figure*}
\subsection{Effective \textit{\textbf{k.p}} Hamiltonian and spin textures} 
To better understand the spin splitting properties, we have constructed the 2-band \textit{\textbf{k.p}} model using the method of invariants, which describes the dispersion around $\Gamma$, M and K points. The minimal \textit{\textbf{k.p}} model is derived by including the spin degrees of freedom. Therefore, we have chosen  eigenstates of Pauli matrix $\sigma_z$, $|\uparrow\rangle$ and  $|\downarrow\rangle$  with eigenvalues +1 and -1, respectively, as the basis set. The little group of $\Gamma$ point is $D_{3h}$. The symmetry allowed Hamiltonian considering the terms up to third order in momentum reads as~\cite{sheoran2022full} (see section III of SM)
\begin{equation}
	H_\Gamma(\textbf{\textit{k}})=H_0(\textbf{\textit{k}})+\lambda k_x(3k_y^2-k_x^2)\sigma_z
	\label{e1}
\end{equation}
where $H_0(\textbf{\textit{k}})$ is the free part of the Hamiltonian having energy eigenvalues $E_0(\textbf{\textit{k}})=\frac{\hbar^2k_x^2}{2m_x^*}+\frac{\hbar^2k_y^2}{2m_y^*}$. Here, $m_x^*$ and $m_y^*$ are the effective masses along $k_x$ and $k_y$ directions, respectively. The lowest order symmetry allowed splitting term is cubic in \textbf{\textit{k}}. Around $\Gamma$ point, bands split along $k_x$ ($\Gamma-$K) direction and doubly degenerate along $k_y$ ($\Gamma-$M) direction. On the other hand, the little group of M point is $C_s$, containing the $\sigma_h$ and $\sigma_v$ operations. The effective \textit{\textbf{k.p}} model for band dispersion around M point is given by (see section III of SM)
\begin{equation}
	H_\textrm{M}(\textbf{\textit{k}})=H_0(\textbf{\textit{k}}) + \alpha k_x \sigma_z
	\label{e3}
\end{equation}
The splitting is linear in \textbf{\textit{k}} around M point and can be treated as complimentary to linear Rashba or linear Dresselhaus. In addition, the splitting is strongly anisotropic in nature i.e., bands are degenerate along $k_y$ (M$-\Gamma$) direction and are lifted along $k_x$ (M-K) direction.  At the K/K$^\prime$ point, associated little group is $C_{3h}$. Thus, \textit{\textbf{k.p}} model is given by (see section III of SM)
\begin{equation}
	\begin{split}
	H_\textrm{K}(\textbf{\textit{k}})=H_0(\textbf{\textit{k}})+k_x(3k_y^2-k_x^2)\;\;\;\;\;\;\;\;\;\;\;\;\;\;\;\;\;\;\;\;\;\;\;\; \\
	+\sigma_z[\Delta +\eta(k_x^2+k_y^2)+\zeta k_x(3k_y^2-k_x^2)]
	\end{split}
\end{equation}
At K/K$^\prime$ point, Zeeman-type splitting in the valence band is predicted by the non-zero coefficient $\Delta$ in  $H_K(\textbf{\textit{k}})$. The $\Delta$ observed for time-reversal conjugate valleys K/K$^\prime$ is same, whereas, the nature of spin splitting is opposite in nature. The properties which are odd under the time-reversal will have opposite nature at K and K$^\prime$ valleys. The spin polarization is one such feature (see Fig.~\ref{p2}(d)). Hence, the spin degrees of freedom and valleys are distinctly locked at inequivalent valleys. Therefore, the flip of only one binary index is forbidden. This strong coupling between spin and valley may enable the long-life valley and spin relaxation.

Figure 4(a) and 4(b) show the spin textures computed using DFT and \textit{\textbf{k.p}} models, respectively. It is clearly seen that in-plane spin components are absent and band dispersion is fully characterized by the out-of-plane spin component. The spin polarization of Bloch states is either parallel or anti-parallel to $z$-direction leading to the case of PST. These spin textures are different from widely reported locally existing persistent spin textures and are preserved in the whole BZ~\cite{jin2021enhanced,jia2020persistent,bhumla2021origin,D1MA00912E}. Therefore, the control of Fermi level to specific part of BZ where PST occurs is not required. These full-zone persistent spin textures (FZPST) can be explained using the symmetry arguments. The little group of the arbitrary $k$-point is $C_s$ containing the $\sigma_h$ operation, besides the trivial identity. This imparts a condition on the spin expectation values at an arbitrary \textbf{\textit{k}}-point as follows
\begin{equation}
	\sigma_h^{-1}s(\textbf{\textit{k}})=s(\textbf{\textit{k}})
\end{equation}
and therefore
\begin{equation}
	(S_x, S_y, S_z)=(-S_x, -S_y, S_z)
\end{equation}
This condition is satisfied only when the in-plane spin components are zero, leading to FZPST along the $z$-direction. This is in accordance with DFT and \textit{\textbf{k.p}} model predictions. The SOC part of Hamiltonian can be written as, $H_{\textrm{SOC}}(\textbf{\textit{k}})=\Omega(\textbf{\textit{k}}).\sigma$, where $\Omega(\textbf{\textit{k}})$ is the spin-orbit field (SOF). In the present case, electron motion accompanied by the spin precession around the unidirectional SOF. This leads to a spatially periodic mode of spin polarization, forming a highly stable persistent spin helix (PSH) state~\cite{schliemann2017colloquium}. The SU(2) conservation laws in the PSH state imply that the spin expectation values have an infinite lifetime~\cite{bernevig2006exact}. Since $S_z$ is a conserved quantity, the fluctuation in the $z$-component of spin polarization with wave vector can only decay by diffusion, thus protecting the spins from dephasing. Figure~\ref{p4}(c) shows the schematic diagram of PSH, where $L_{\textrm{PSH}}$ is the wavelength of PSH. The angular frequency ($\omega$) of precession around SOF can be expressed as $\omega=-\gamma B$, where $\gamma$ is the gyromagnetic ratio. B is the effective magnetic field coming from Eq.~\ref{e3} and can be written as $B=2\alpha k_x/\gamma\hbar$. The spin precession angle ($\theta$) around the $y$-axis for an electron moving in real space, under influence of unidirectional SOF at the time t is, $\theta=\omega t=2\alpha k_x t/\hbar$. The time at which precession completes a revolution ($\theta=2\pi$) is, $t=\pi\hbar/\alpha k_x$. The distance traveled by the electron along $y$-axis in this time period is, $L_{\textrm{PSH}}=vt=\hbar k_x t/m^*$. Therefore, the PSH length is given by 
\begin{equation}
	L_{\textrm{PSH}}=\frac{\pi \hbar^2}{m^* \alpha}
	\label{e7}
\end{equation}
The length of PSH mode can be calculated by evaluating the band dispersion parameters $m^*$ and $\alpha$. The PSH mechanism leads to the long-range spin transport without dissipation, enabling better efficiency in spintronics devices.
\begin{table}
	\begin{center}
		\caption{The calculated SOC parameters, viz. $\alpha$, $\Delta$ and wavelength of PSH mode ($L_{\textrm{PSH}}$) of upper valence bands for WA$_2$Z$_4$ family. }
		\label{t2}
		\begin{tabular}{p{2cm} p{1.5cm} p{1.5cm} p{1.5cm} p{1.5cm} }
			\hline
			\hline
			& phase   & $\alpha$ & $\Delta$ & $L_{\textrm{PSH}}$   \\
			&  & (eV{\AA}) & (eV) & (nm)  \\\hline  
			WSi$_2$N$_4$ & $\alpha_1$  & 0.87 & 0.384 & 5.03  \\
			& $\alpha_2$  &0.92 & 0.397 & 4.84  \\ 
			WGe$_2$N$_4$ & $\alpha_1$ & 1.20 & 0.435 & 4.51  \\
			& $\alpha_2$ & 1.21& 0.437 &  4.49  \\ 
			WSi$_2$P$_4$ & $\alpha_1$ & 1.19  & 0.407 & 4.42  \\
			& $\alpha_2$ & 1.22 & 0.429 &  4.31 \\ 
			WGe$_2$P$_4$ & $\alpha_1$ & 1.37 & 0.450 & 4.08 \\
			& $\alpha_2$ & 1.33 & 0.442 & 4.07  \\ 
			WSi$_2$As$_4$& $\alpha_1$ & 1.47 & 0.459 & 3.99  \\
			& $\alpha_2$ & 1.61 & 0.496 & 3.21 \\ 
			WGe$_2$As$_4$& $\alpha_1$ & 1.74 & 0.499 & 3.16   \\
			& $\alpha_2$ & 1.79 & 0.507 &  2.89 \\ 
			\hline\hline
		\end{tabular}
	\end{center}
\end{table}

In order to quantify the strength of splitting effects, we parameterize the splitting coefficients ($\lambda$, $\alpha$, $\Delta$, $\eta$ and $\zeta$) by fitting the DFT band structures. In general, we are interested in coefficients having the largest contribution in band splitting, i.e., $\lambda$, $\alpha$,  and $\Delta$. Figure~\ref{p4}(d)-(f) show the comparison between the DFT and model resulted band splits. The splitting observed around $\Gamma$ point is very tiny ($\lambda=0.21$ eV{\AA}$^3$), as the energy contribution coming from the term cubic in \textbf{\textit{k}} is very small i.e., $k_x\ll1$ {\AA}. The linear splitting observed around M point is $\alpha=0.92$ eV{\AA}. The energy eigenvalues of \textit{\textbf{k.p}} model given in Eq.~\ref{e3} can be written as
\begin{equation}
	E_\textrm{M}^{\pm}=\frac{\hbar^2}{2m^*}({|k|\pm k_0})^2+E_R
\end{equation}
where $E_R$ and $k_0$ are the shifting energy and wave vector, respectively, calculated from the spin-split bands along the $k_x$ ($\Gamma-$K) direction as shown in Fig.~\ref{p4}(e). The $\alpha$ can also be estimated using $E_R$ and $k_0$ as follows
\begin{equation}
	\alpha=\frac{2E_R}{k_0}
\end{equation}
The values of $E_R = 35$ meV and $k_0=0.075$ {\AA}, lead to $\alpha=0.93$ eV{\AA}. In addition, we calculate the $L_{\textrm{PSH}}$ from the Eq.~\ref{e7} and it is found to be  $4.84$ nm for $\alpha_2$-WSi$_2$N$_4$. Table~\ref{t2} shows the calculated values of  $\alpha$, $\Delta$ and $L_{\textrm{PSH}}$ for the WA$_2$Z$_4$ family. The effect of SOC increases with increasing atomic weight, therefore, the splitting is in the order Ge$>$Si and As$>$P$>$N. The observed values of $\alpha$ are larger than traditional TMDs ($0.05$ - $0.40$ eV{\AA})~\cite{hu2018intrinsic}, WO$_2$Cl$_2$ monolayer ($0.90$ eV{\AA})~\cite{ai2019reversible}, defective PtSe$_2$ (0.20 - 1.14 eV{\AA})~\cite{absor2017defect} and (Mo,W)X$_2$ (X=S,Se)(0.14 - 0.26 eV{\AA})~\cite{li2019unidirectional}.

\begin{figure*}
	\includegraphics[width=16cm]{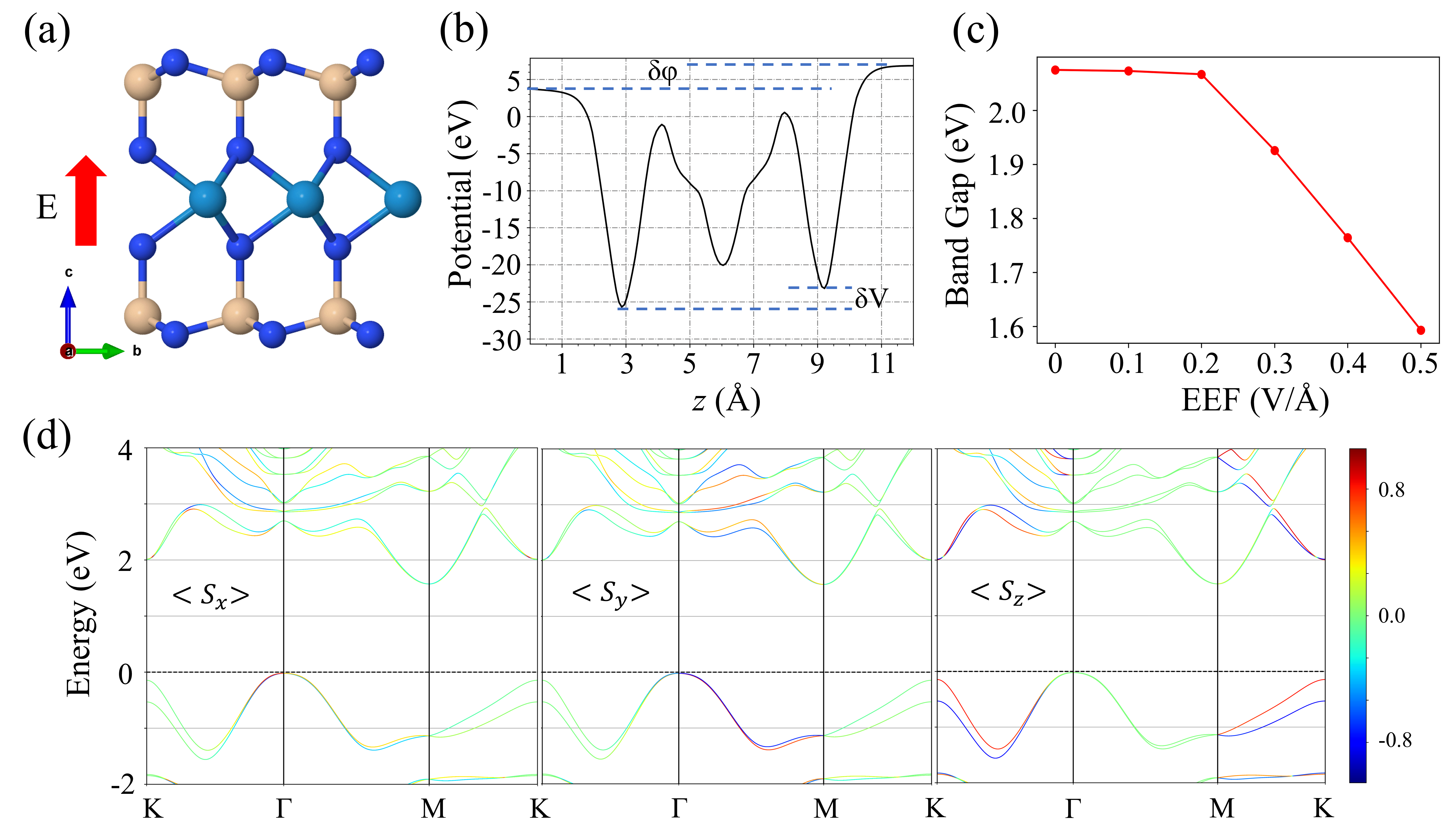}
	\caption{ The direction of EEF shown with respect to the side view of $\alpha_2$-WSi$_2$N$_4$. (b) The planar average electrostatic potential under the application of EEF as a function of out-of-plane axis. Here, $\delta$V denotes the electrostatic potential difference between two Si layers and $\delta \phi$ denotes the workfunction difference between two outer surfaces. (c) Variation of the band gap as a function of EEF. (d) The spin-projected band structure under the EEF of strength 0.5 V{\AA}. The color bar shows the $x$-, $y$- and $z$- components of spin polarization.}
	\label{p5}
\end{figure*}

\subsection{Impact of electric field}
\begin{figure}[h]
	     \includegraphics[width=8.5cm]{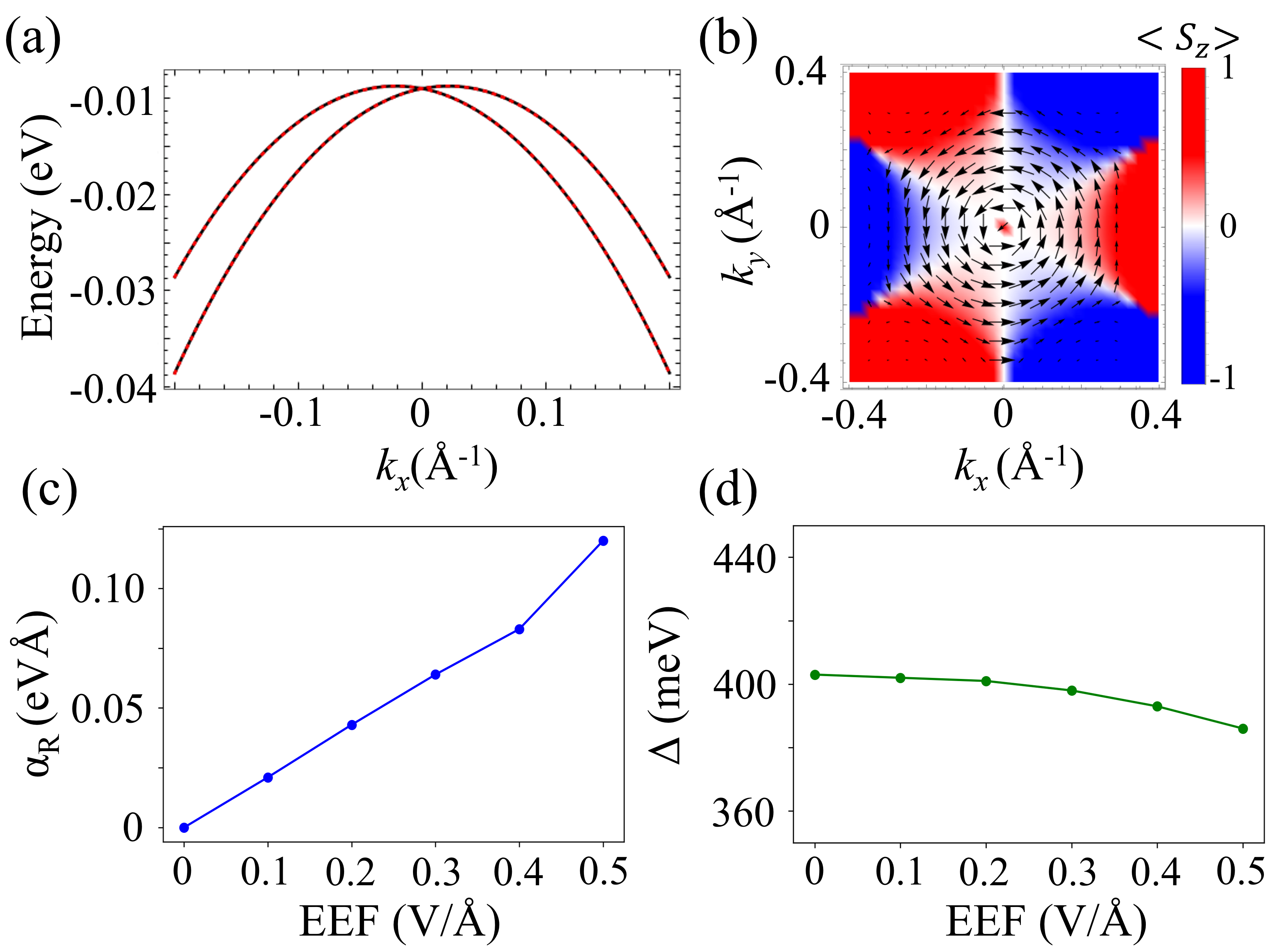}
	     \caption{(a) The topmost valence bands and the associated (b) spin texture around the $\Gamma$ point for of $\alpha_2$-WSi$_2$N$_4$ monolayer. The black solid lines and red dots represent the band structure computed using DFT and \textit{\textbf{k.p}} model, respectively. The arrows and color bar show the in-plane ($x$- and $y$-) and out-of-plane ($z$-) components of spin texture, respectively. The variation of (c) $\alpha_R$ and (d) $\Delta$ as a function of EEF.}
	     \label{p6}
\end{figure}

An external electric field (EEF) is always considered as an effective approach to manipulate and introduce new splittings by breaking the inversion symmetry, without changing the time-reversal invariant symmetry. Previous studies for TMDs have shown that the strength of Rashba-type splitting is tunable with the help of EEF~\cite{hu2018intrinsic}. To understand the effect of EEF, we theoretically apply the EEF in the range of 0 to 0.5 V/{\AA}, in the $z$-direction on $\alpha_2-$WSi$_2$N$_2$, as illustrated in Fig.~\ref{p5}(a). Firstly, we plot the planar average of the electrostatic potential energy as a function of $z$. In contrast to the case of without EEF, the potential energy here is asymmetric with respect to the W atom (see Fig.~\ref{p5}(b)). It can be easily seen that a significant potential difference ($\delta V=2.3$ eV) is generated between the two Si-layers. The potential energy on the upper side of Si layer is larger compared to its lower counterpart. This confirms the breaking of the surface inversion symmetry. Then, we observe the effect of EEF on band gaps, which can be explained as stark effect in condensed matter system. Our system is out-of-plane symmetric, therefore, EEF shows the same effect in both positive and negative $z$-directions. Figure~\ref{p5}(c) shows the band gap variation as a function of EEF. Between 0.1 and 0.2 V/{\AA}, the position of CBm changes from K point to the M point (see Fig. S5 in SM). The band gap is nearly constant up to 0.2 V/{\AA} and thereafter, decreases linearly to 1.59 eV under 0.5 V/{\AA} EEF. Figure~\ref{p5}(d) shows the spin projected band structure with an EEF of 0.5 V/{\AA}. Here, we see that near VBM ($\Gamma$ point), the out-of-plane spin component vanishes and the in-plane spin components arise. Figure~\ref{p6}(a) and~\ref{p6}(b) show the band structure around $\Gamma$ point and associated spin texture, respectively. The helical-type in-plane spin polarization confirms the existence of Rashba-type splitting. 
\begin{figure*}
	\includegraphics[width=14cm]{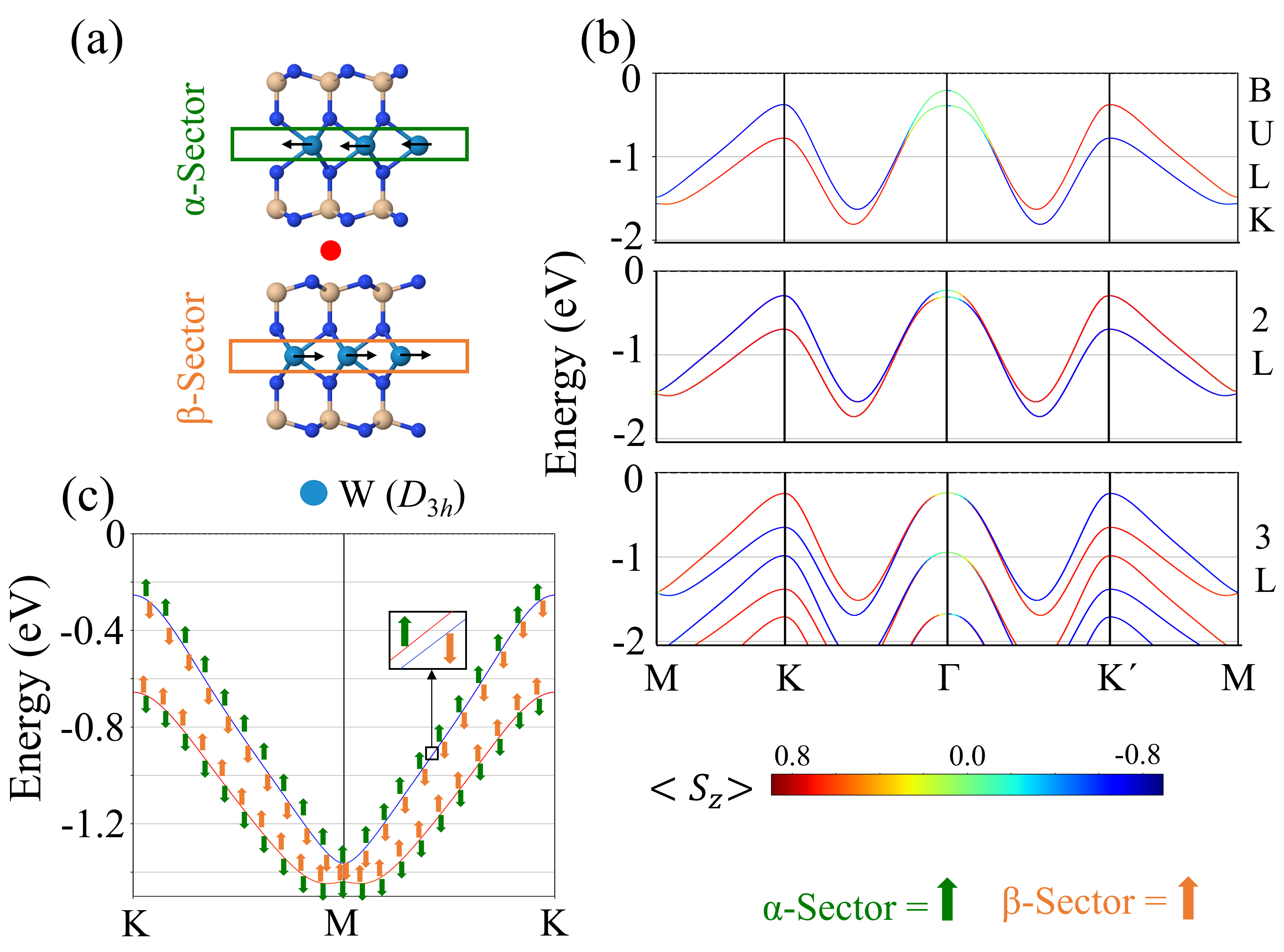}
	\caption{ (a) Side view of the bulk crystal structure of WSi$_2$N$_4$ with site point group of W atom. The red dot represents the inversion center. The two boxed real-space sectors forming the inversion partners are used for spin projection are considered as $\alpha$-sector and $\beta$-sector. The black arrows on W atoms represent the direction of dipole moment on particular atomic sites. (b) The spin-projected top valence bands of bulk WSi$_2$N$_4$, bilayer (2L) and trilayer (3L). The color bar shows the out-of-plane ($z$-) component  of spin textures. (c) The spin- and atom- projected band structure of top valence bands along K-M-K path. The green and orange color shows the band is coming from $\alpha$- and $\beta$-sectors, respectively. The up and down arrows shows the $z$-component of spin polarization. Note that the in-plane spin components are zero along that path.}
	\label{p7}
\end{figure*}

To understand Rashba splitting, we include additional EEF dependent term in the aforementioned \textit{\textbf{k.p}} models, which is given as
\begin{equation}
	H^{\textrm{EEF}}(\textbf{\textit{k}})=\alpha_R(\sigma_xk_y-\sigma_yk_x)
	\label{eeef}
\end{equation}
where $\alpha_R$ is the splitting coefficient dependent on the strength of EEF. The  $H^{\textrm{EEF}}(\textbf{\textit{k}})$ term breaks the SIS, which reduces the point group symmetry of 2D system from $D_{3h}$ to $C_{3v}$ and considers the in-plane spin component. The contribution of $H^{\textrm{EEF}}(\textbf{\textit{k}})$ near $\Gamma$ point is dominant as compared to cubic term, therefore, the in-plane spin component is observed near $\Gamma$ point. Furthermore, the out-of-plane component starts dominating as we move away from the $\Gamma$ point. Figure~\ref{p6}(c) and~\ref{p6}(d) show the variation of $\alpha_R$ (around $\Gamma$ point) and $\Delta$ as a function of EEF. The $\alpha_R$ is linearly dependent on the EEF. Such a linear relationship helps in the precise control of spin-precession in spin field-effect transistors. When the EEF is 0.5 V/{\AA}, the Rashba spin splitting occurs at the $\Gamma$ point with a Rashba coefficient of 0.12 eV{\AA}, obtained by parameterizing the Eq.~\ref{eeef}. The observed values of $\alpha_R$ are small as compared to the TMD ($\alpha_R=$0.20 eV{\AA}) under the same EEF, since the W atom is covered by the Si-N layers. Thus, the W-$d$ orbitals contributing the VBM are less influenced by the EEF. The spin splitting and spin textures around K and M points are hardly influenced by the EEF, because the non-zero constant term $\Delta$ around K and linear term with coefficient $\alpha$ dominate over the $H^{\textrm{EEF}}(\textbf{\textit{k}})$. Therefore, the SOFs $\Omega_K(\textbf{\textit{k}})$ and $\Omega_M(\textbf{\textit{k}})$ are parallel to $z$-direction and are independent of EEF. The PST preserve fully along M-K path having only the out-of-plane component of spin polarization. Therefore, PST are robust to small EEF, except around $\Gamma$ point.

\subsection{Hidden spin polarization in bulk WSi$_2$N$_4$}
The previous theoretical study reports the possibility of experimental realization of layered bulk WSi$_2$N$_4$ family, using the bottom up synthesis method~\cite{islam2021tunable}. We have considered the layer dependence of the spin polarization in layered WSi$_2$N$_4$ family.  The bulk WSi$_2$N$_4$ has an indirect band gap of 1.97 eV having CBm and VBM at K and $\Gamma$, respectively~\cite{islam2021tunable}. Figure~\ref{p7}(a) shows the unit cell of bulk WSi$_2$N$_4$ having $P6_3/mmc$ space group symmetry. The associated point group symmetry is $D_{6h}$ and have a point of inversion as shown in Fig.~\ref{p7}(a). The presence of inversion and time reversal symmetries lead to the bands to be two-fold degenerate throughout the BZ. As expected, the top valence bands are two-fold degenerate as shown in Fig.~\ref{p7}(b). In current view, the Rashba and Dresselhaus effects are observed only when the inversion symmetry is absent. Thus, it is unwise to look for spin splitting and spin polarization in centrosymmetric bulk WSi$_2$N$_4$. Therefore, the degree of spin polarization is expected to be zero in inversion symmetric even layered and bulk WSi$_2$N$_4$, as reported in Ref.~\cite{islam2021tunable}. In contrast, we report the existence of hidden spin polarization for bulk WSi$_2$N$_4$, which can be attributed to the hidden Dresselhaus (D-2) effect~\cite{zhang2014hidden}. The key point here is that SOC is anchored on particular nuclear sites in the solid, thus it is the individual atomic site symmetry in solid that forms a good starting point to describe the SOC-induced spin-polarization. The net spin polarization of the bulk crystal is sum of the local polarization over the atomic sites. The D-2 effect arises when the site point group of an atom within a 3D crystal lacks inversion asymmetry and the crystal as a whole is centrosymmetric~\cite{zhang2014hidden}. 

In present case, the inversion symmetry is present in bulk space group but not in the site point groups (see Fig.~\ref{p7}(a)). One such atomic site is W atom, where site point group is inversion asymmetric $D_{3h}$, contributing mainly to the low-energy spectrum. The two W layers connected by the inversion symmetry are indicated as sectors $S_\alpha$ and $S_\beta$. The two sectors are individually inversion asymmetric and produce equal dipole moment but in opposite directions (see Fig.~\ref{p7}(a)). Therefore, the net dipole moment of bulk WSi$_2$N$_4$ becomes zero, which is the central property of centrosymmetric structures. Figure~\ref{p7}(c) shows the band structure projected on the spin components and real space sectors $S_\alpha$ and $S_\beta$. If we consider only one sector, the large band splitting ($\sim$0.4 eV) between spin-up and spin-down states is observed along K-M-K path. The interactions between the sectors form a pair of bonding and antibonding states, therefore, open a band gap at the band crossing M point~\cite{liu2015search}. Even though the bands are doubly degenerate as a consequence of bulk inversion asymmetry, the branches of doubly degenerate bands have opposite spin polarizations, which are separated in the real space (see Fig.~\ref{p7}(c)). As already discussed, $S_z$ is a good quantum number, therefore, the spins are fully out-of-plane and up and down spins locate at the energy bands. The weak van der Waals interaction among inversion partners causes slight spin mixture from another layer and  slightly larger separation of the degenerate bands. However, most of the spin at the K valley remains on each layer. The real space segregation of spin polarized bands originates from the separate local sites Dresselhaus SOC. The bulk spin polarization in such systems can be observed by controlling the stacking directions probing beams~\cite{riley2014direct, yao2017direct, wu2017direct}. The extra advantage of D-2 effect is that the spins can be controlled by manipulating the interaction between two inversion parts. One such example is that a small electric field ($\sim$$1$ mV/{\AA}) is required to reverse the spin polarization as compared to linear Rashba effect~\cite{yuan2019uncovering}. 

The results similar to the bulk are expected to be observed for even number of layers of WSi$_2$N$_4$, having centrosymmetric $P6_3/mmc$ space group symmetry. Therefore, we have examined the results for bilayer-WSi$_2$N$_4$. The real space segregated hidden spin polarization is also observed in doubly degenerate bands of topmost valence bands (see Fig.\ref{p7}(b)). The band splitting similar to monolayer is observed in odd number of layers of WSi$_2$N$_4$ having non-centrosymmetric $P\overline{6}m2$ space group symmetry (i.e. see band structure for trilyer in Fig.\ref{p7}(b)).

Now, let us comment on the possible applications of PST in WA$_2$Z$_4$ materials, which can act as channel for spin-FET. The out-of-plane spin polarization is injected through the ferromagnetic source. The SOC rotate its polarization during its path, and is either transmitted or reflected to the drain electrode depending upon the spin state, therefore, act as on or off states, respectively~\cite{schliemann2017colloquium}. The on/off functionality can be further controlled by the EEF. The observed small value of $L_{\textrm{PSH}},  i.e.,2.89$ nm is promising for highly efficient devices. The coexistence of hidden spin polarization and PST is another important avenue to pursue. 

\section{Conclusion}
We have revealed the interesting spin splitting properties of WA$_2$Z$_4$ family. The formation energies, phonon spectra and AIMD calculations confirm the stability of these materials. The PBE, HSE06 and the excited state methods G$_0$W$_0$ and BSE show that these materials are semiconducting, having a band gap in visible region. In addition to valley dependent properties at the corners of BZ, we have observed cubic and linear split bands around the time-reversal invariant $\Gamma$ and M points, respectively. The in-plane mirror symmetry ($\sigma_h$) leads to out-of-plane FZPST, therefore, enabling dissipationless long range transport through spatially periodic mode PSH mechanism. The effect of EEF is duly considered on band gap and spin splitting. The EEF leads to linear Rashba-type helical spin texture around the VBM. The impact of layer thickness is studied in aforementioned effects. We have found the hidden spin polarization in the bulk and even layered WSi$_2$N$_4$, which can be attributed to the D-2 effect. The D-2 effect arises because the local site point of W atom lacks inversion symmetry; however, the global space group is centrosymmetric. The PST is robust to electric field and thickness along the M-K direction. We believe that this work will provide highly efficient spintronics and valleytronics devices for room temperature applications. 

\section{Acknowledgement}
S.S. acknowledges CSIR, India, for the senior research fellowship [grant no. 09/086(1432)/2019-EMR-I]. D.G. acknowledges UGC, India, for the senior research fellowship [Grant No. 1268/(CSIR-UGC NET JUNE 2018)]. A.P. acknowledges IIT Delhi for the junior research fellowship. S. B. acknowledges financial support from SERB under a core research grant (grant no. CRG/2019/000647) to set up his High Performance Computing (HPC) facility ``Veena" at IIT Delhi for computational resources. We acknowledge the High Performance Computing (HPC) facility at IIT Delhi for computational resources.

\bibliography{references}

\end{document}


\title{Supplemental Material for\\
	``Coupled spin-valley, Rashba effect and hidden persistent spin polarization in WSi$_2$N$_4$ family"}
\author{Sajjan Sheoran\footnote{sajjan@physics.iitd.ac.in}, Deepika Gill, Ankita Phutela, Saswata Bhattacharya\footnote{saswata@physics.iitd.ac.in}} 
\affiliation{Department of Physics, Indian Institute of Technology Delhi, New Delhi 110016, India}
\pacs{}
\keywords{DFT, \textbf{\textit{k.p}} method, Valley properties, Rashba effect, Hidden spin polarization, WSi$_2$N$_4$ family}
\maketitle
\begin{enumerate}[\bf I.]
	
	\item Electronic structure of WA$_2$Z$_4$ using PBE+SOC and HSE06+SOC
	\item Electronic and optical properties of WSi$_2$N$_4$ using beyond DFT methods
	\item The \textbf{\textit{k.p}} models
	\item Electric field effect on electronic structure
	
\end{enumerate}
\section{Electronic structure of WA$_2$Z$_4$ using PBE+SOC and HSE06+SOC}

\begin{figure}[H]
	\includegraphics[width=18cm]{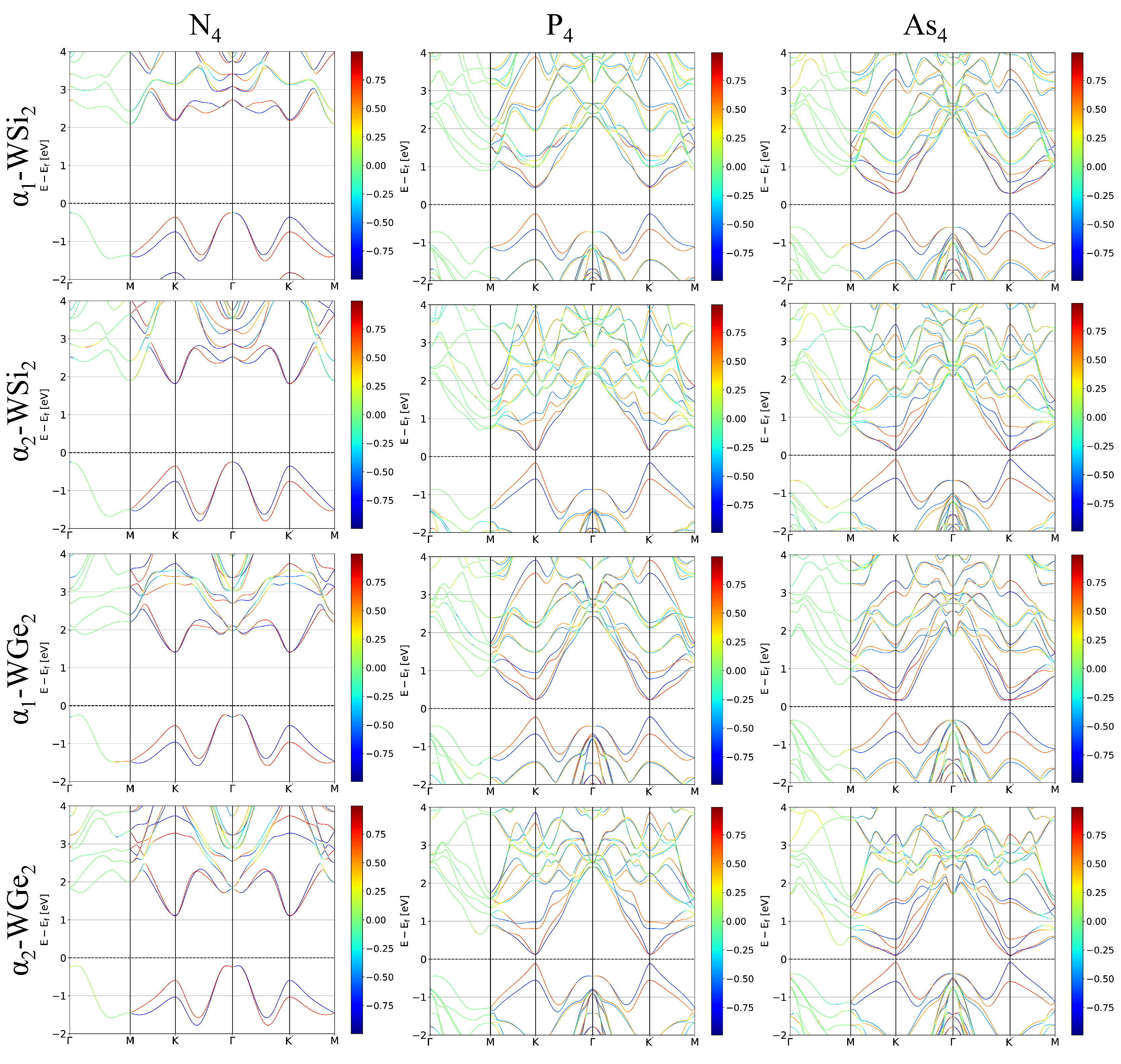}
	\caption{ The calculated band structure of $\alpha_1$- and $\alpha_2$-WA$_2$Z$_4$ (A=Si, Ge; Z=N, P, As) using PBE+SOC $\epsilon_{xc}$ functional. The color bars show the $z$-components of spin polarization.}
\end{figure}
\begin{figure}[H]
	\includegraphics[width=18cm]{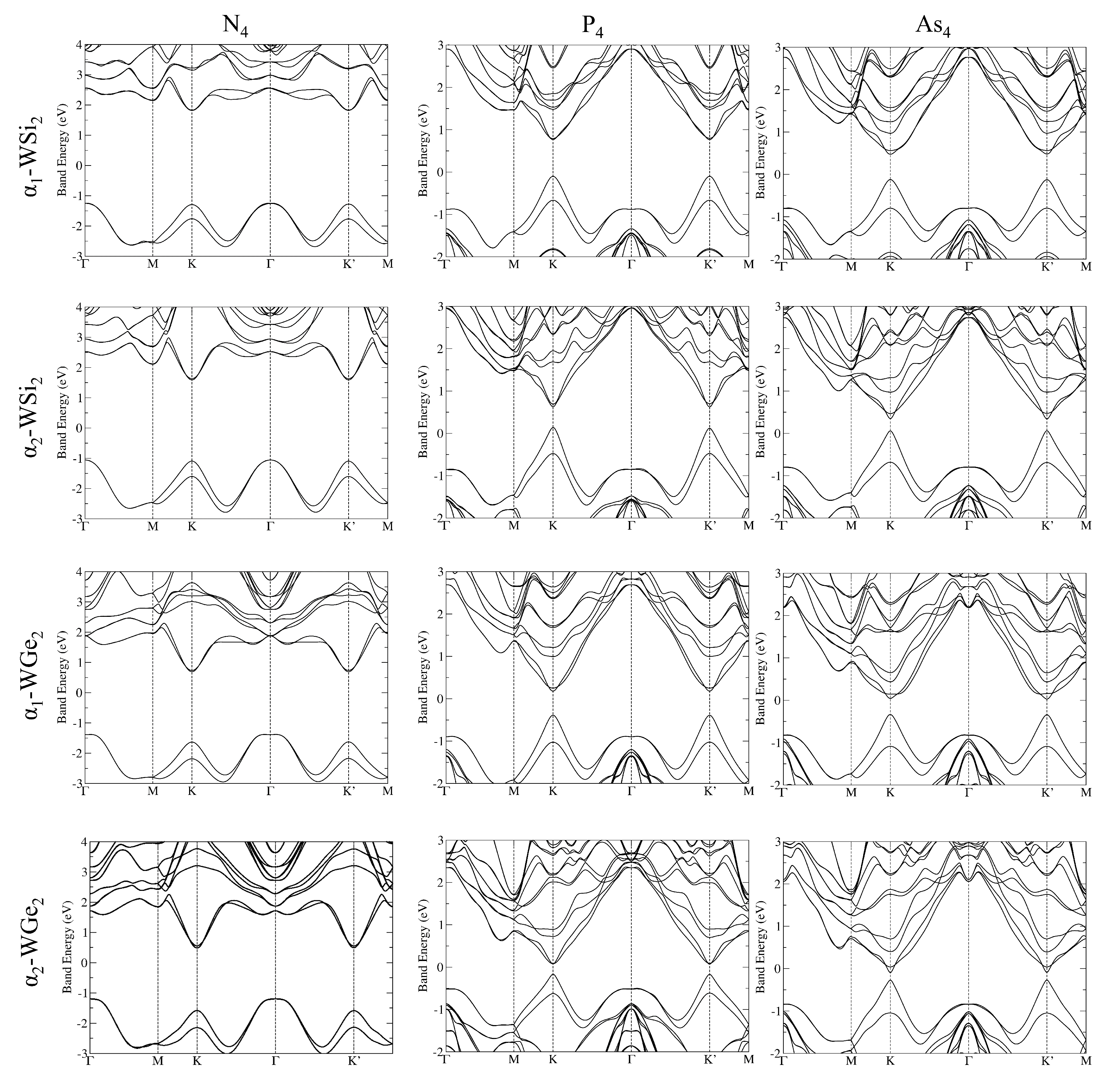}
	\caption{ The calculated band structure of $\alpha_1$- and $\alpha_2$-WA$_2$Z$_4$ (A=Si, Ge; Z=N, P, As) using HSE06+SOC $\epsilon_{xc}$ functional.}
\end{figure}
\newpage
\section{Electronic and optical properties of WSi$_2$N$_4$ using beyond DFT methods}
\begin{figure}[H]
	\begin{center}
	\includegraphics[width=8cm]{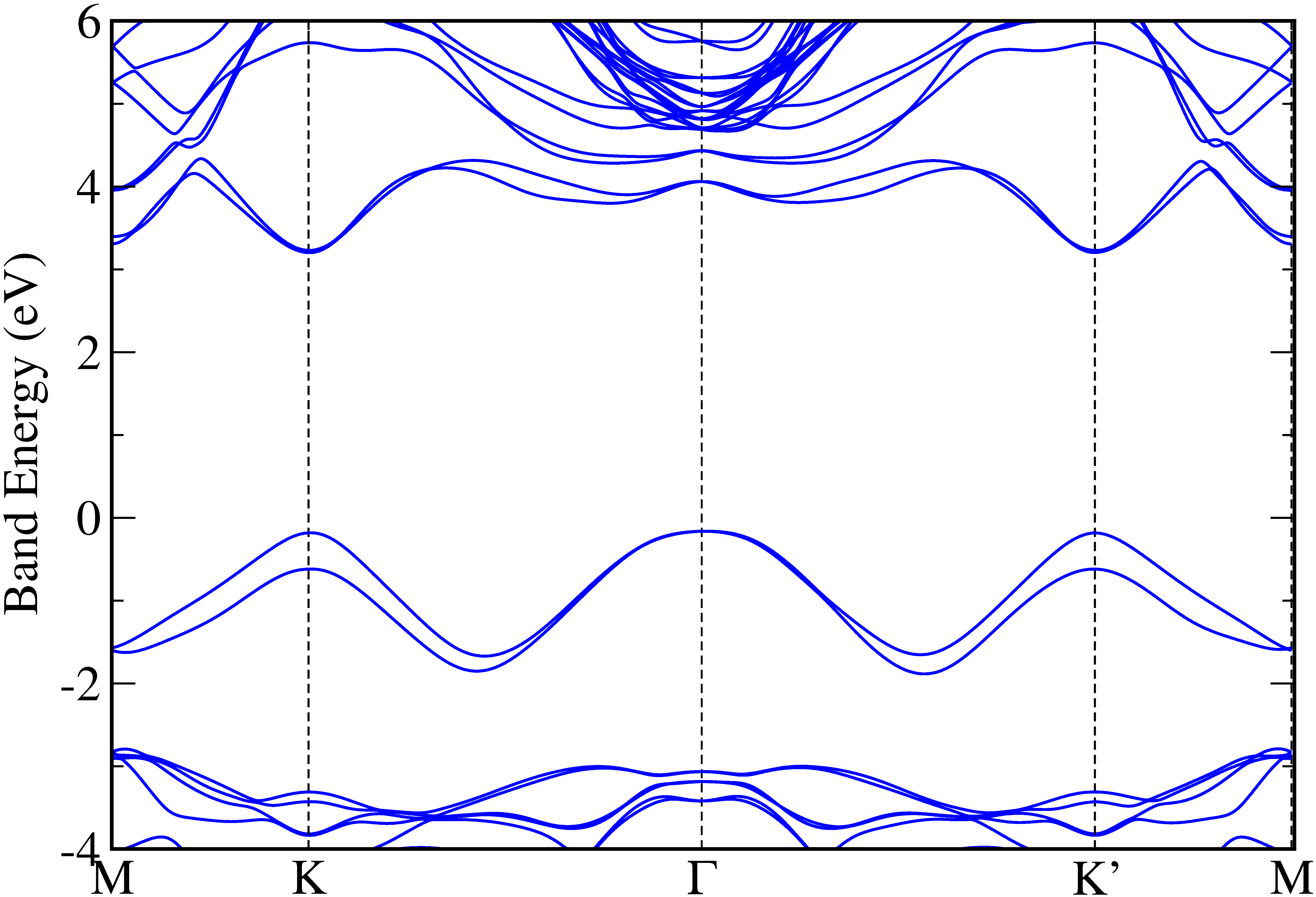}
	\caption{ The calculated band structure of $\alpha_2$-WSi$_2$N$_4$ using G$_0$W$_0$ performed on the top of PBE+SOC (G$_0$W$_0$@PBE+SOC).}
	\label{S3}
    \end{center}
\end{figure}

\begin{figure}[H]
	\includegraphics[width=16cm]{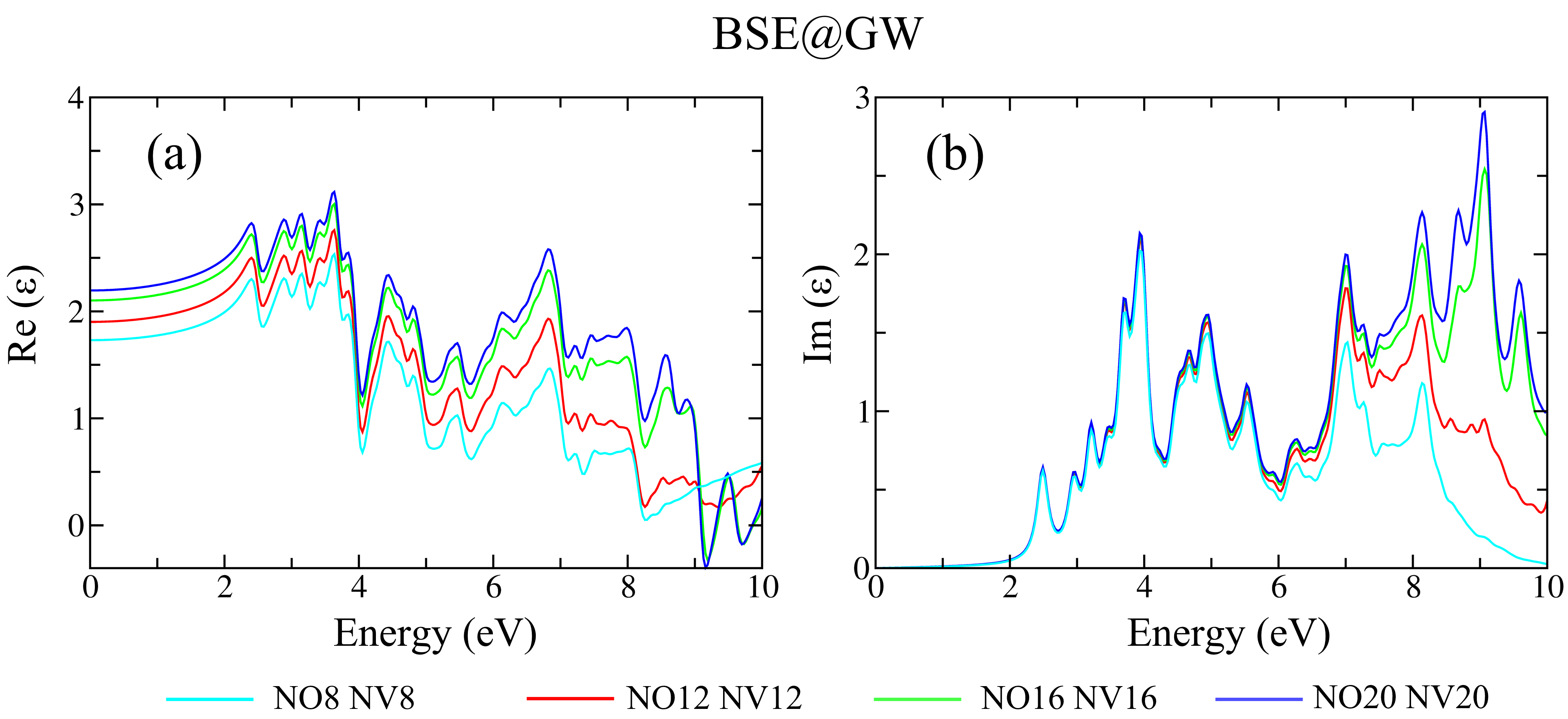}
	\caption{ The calculated (a) real and (b) imaginary parts of dielectric function for $\alpha_2$-WSi$_2$N$_4$. Here, NO and NV signify the numbers of occupied and nonoccupied orbitals taken into consideration, respectively.}
	\label{S4}
\end{figure}

It is well understood that the Hohenberg-Kohn density functional theory (DFT) is strictly limited to the ground state properties. Therefore, we study the optical properties using DFT and beyond approaches under the framework of many body perturbation theory, i.e. G$_0$W$_0$ and BSE. The G$_0$W$_0$ approach takes the screened coulombic interaction into consideration, using the perturbative method and improves on the Hartree-Fock approximation. In our case we perform the single shot G$_0$W$_0$ calculations on the top of the orbitals obtained from the PBE calculations. Firstly, we converge the band gap with respect to the k-grid, number of bands and energy cutoff. We take number of bands (NBANDS) to be four times the number of occupied orbitals. We take \textbf{\textit{k}}-grid, NBANDS and energy cutoff to be 10$\times$10$\times$1, 240 and 550 eV, respectively. The calculated band gaps for WA$_2$Z$_4$ is given in the Table 1 of the main text. Fig.~\ref{S3} shows the band structure calculated using the G$_0$W$_0$@PBE+SOC for $\alpha_2$-WSi$_2$N$_4$. We have obtained the indirect band gap of 3.36 eV. Here, we see that apart from the band gap, the band dispersion curves remains similar to the PBE+SOC and HSE06+SOC. We calculate the excitonic effect using the BSE approach, which is second order Green's function technique. The Fig.~\ref{S4} shows the real (Re ($\epsilon$)) and imaginary (Im ($\epsilon$)) parts of the dielectric function. The Re ($\epsilon$) and Im ($\epsilon$) are thoroughly verified by considering the different number of occupied and unoccupied bands. First excitonic peak obtained using BSE is at 2.49 eV confirming that $\alpha_2-$WSi$_2$N$_4$ is sensitive to visible spectrum. The exciton binding energy ($E_B$) can be computed from the difference between the quasiparticle band gap (G$_0$W$_0$ band gap) and optical band gap (first BSE peak). For $\alpha_2$-WSi$_2$N$_4$, exciton binding energy is 0.87 eV. The $E_B$  observed is comparable to monolayer transition metal dichalcogenides (0.6 eV-1.0 eV), owing to strong coulomb interaction~\cite{mueller2018exciton}. The observed $E_B$ is one-two order magnitude larger than the conventional GaAs~\cite{chernikov2014exciton, andrews2019comprehensive}. Therefore, excitonic features are stable at room temperature and dominates the optical response and non equilibrium dynamics of these materials. The large $E_B$ induces strong excitonic effects, i.e., the large oscillator strength induces strong light matter interaction and absorbance as high as 0.1-0.3~\cite{andrews2019comprehensive}.

\section{The \textbf{\textit{k.p}} models}
\subsection{$\Gamma$ point}
\begin{table}[h]
	\begin{center}
		\caption {The transformations of ($\sigma_x$, $\sigma_y$, $\sigma_z$) and ($k_x$, $k_y$, $k_z$) with respect to the generators of the $D_{3h}$ point group and time-reversal operator ($T$). The first row shows the point group operations. Note that the generators are enough to form the whole group. Hence, only these generators along with time-reversal operation $T=i\sigma_yK$ ($K$ is complex conjugation operator) are considered to construct the \textbf{\textit{k.p}} model. The last row shows the terms which are invariant under point group operation. Note that we have included the terms upto cubic in \textbf{\textit{k}} and higher order contributions are found to be insignificant.}
		\begin{tabular}{p{1.8cm}p{3.0cm}p{3.8cm}p{3.8cm}p{2.cm}}
			\hline
			\hline
			& $C_{3z}=e^{{-i\pi}/{3\sigma_z}}$ & $M_{yz}=i\sigma_x$ & $M_{xy}=i\sigma_z$& $T=i\sigma_yK$           \\ \hline\hline
			
			$\;\;\;\,k_x$       &  $-k_x/2+\sqrt{3}k_y/2$  &$\;-k_x$ & $\;\;\;k_x$& $-k_x$        \\ 
			$\;\;\;\,k_y$       &  $-\sqrt{3}k_x/2-k_y/2$  & $\;\;\;\,k_y$ &$\;\;\;k_y$& $-k_y$      \\ 
			$\;\;\;\,k_z$       &  $\;\;\;\,k_z$  & $\;\;\;\,k_z$ &$-k_z$& $-k_z$       \\ 
			$\;\;\;\,\sigma_x$  &  $-\sigma_x/2+\sqrt{3}\sigma_y/2$  & $\;\;\;\,\sigma_x$   &$-\sigma_x$& $-\sigma_x$   \\ 
			$\;\;\;\,\sigma_y$  &  $-\sqrt{3}\sigma_x/2-\sigma_y/2$  & $\;-\sigma_y$  &$-\sigma_y$ & $-\sigma_y$   \\ 
			$\;\;\;\,\sigma_z$  &  $\;\;\;\,\sigma_z$  &$\;-\sigma_z$   &$\;\;\,\sigma_z$& $-\sigma_z$  \\ \hline\hline
			&  $(k_x\sigma_y-k_y\sigma_x$),  & $k^m_ik_x\sigma_y$, $k^m_ik_x\sigma_z$,    & $k^m_ik_x\sigma_z$, $k^m_ik_y\sigma_z$,& $k_i\sigma_j $ \\
			Invariants            &  $k_y(3k_x^2-k_y^2)\sigma_z$,  & $k^m_ik_y\sigma_x$, $k^m_ik_z\sigma_x$   &$k^m_ik_z\sigma_x$, $k^m_ik_z\sigma_y$ & ($i,j=$ \\ 
			& $k_x(k_x^2-3k_y^2)\sigma_z$ & ($i=x,y,z$; $m=0,2$)  & ($i=x,y,z$; $m=0,2$)& $x,y,z$) \\ \hline \hline
		\end{tabular}
		\label{tran1}
	\end{center}
\end{table}
Let us first construct the \textit{\textbf{k.p}} model for the $\Gamma$ point located at the center of Brillouin zone, having the full symmetry of point group of $D_{3h}$. The $D_{3h}$ point group contains identity operation ($E$), three-fold rotation ($C_{3z}$) having $z$ as the principal axis, mirror operations perpendicular to principal axis ($\sigma_h$) and parallel to principal axis ($\sigma_v$). We construct the two-band \textit{\textbf{k.p}} model including the spin degrees of freedom (denoted by the $\sigma_x$, $\sigma_y$, and $\sigma_z$) and \textbf{k} vectors away from the gamma point (denoted by the $k_x$, $k_y$ and $k_z$). Using the approach similar to those employed in Ref.~\cite{voon2009kp}, the transformations of ($\sigma_x$, $\sigma_y$, and $\sigma_z$) and ($k_x$, $k_y$ and $k_z$) under the point group operations and time-reversal operation are given in Table~\ref{tran1}. Therefore the effective Hamiltonian collecting all the symmetry invariants are given by
\begin{equation}
	H_\Gamma(\textbf{\textit{k}})=H_0(\textbf{\textit{k}})+\lambda k_x(3k_y^2-k_x^2)\sigma_z
	\label{e1}
\end{equation}
where, $H_0(\textbf{\textit{k}})$ is the free part of Hamiltonian and $\lambda k_x(3k_y^2-k_x^2)\sigma_z$ is perturbative SOC term.
\subsection{K point}
\begin{table}[h]
	\begin{center}
		\caption {The transformations of ($\sigma_x$, $\sigma_y$, $\sigma_z$) and ($k_x$, $k_y$, $k_z$) with respect to the generators of the $C_{3h}$ point group.}
		\begin{tabular}{p{5cm}p{5cm}p{5cm}}
			\hline
			\hline
			& $C_{3z}=e^{{-i\pi}/{3\sigma_z}}$ & $M_{xy}=i\sigma_z$       \\ \hline\hline
			
			$\;\;\;\,k_x$       &  $-k_x/2+\sqrt{3}k_y/2$  & $\;\;\;\,k_x$ \\ 
			$\;\;\;\,k_y$       &  $-\sqrt{3}k_x/2-k_y/2$  & $\;\;\;\,k_y$\\ 
			$\;\;\;\,k_z$       &  $\;\;\;\,k_z$   &$\;-k_z$       \\ 
			$\;\;\;\,\sigma_x$  &  $-\sigma_x/2+\sqrt{3}\sigma_y/2$  & $\;-\sigma_x$   \\ 
			$\;\;\;\,\sigma_y$  &  $-\sqrt{3}\sigma_x/2-\sigma_y/2$  &$\;-\sigma_y$  \\ 
			$\;\;\;\,\sigma_z$  &  $\;\;\;\,\sigma_z$  &$\;\;\;\,\sigma_z$ \\ \hline\hline
			&  $(k_x\sigma_y-k_y\sigma_x$),  & $k^m_ik_x\sigma_z$, $k^m_ik_y\sigma_z$, \\
			Invariants            &  $k_y(3k_x^2-k_y^2)\sigma_z$,    &$k^m_ik_z\sigma_x$, $k^m_ik_z\sigma_y$  \\ 
			& $k_x(k_x^2-3k_y^2)\sigma_z$  & ($i=x,y,z$; $m=0,2$)\\ \hline \hline
		\end{tabular}
		\label{tran2}
	\end{center}
\end{table}
The little group for the K point is the $C_{3h}$, containing identity operation ($E$), three-fold rotation ($C_{3z}$) having $z$ as the principal axis and mirror operations perpendicular to principal axis ($\sigma_h$). Note that the K point is not the time-reversal invariant point, therefore the Hamiltonian for the K point need not to be time-reversal symmetric. Table~\ref{tran2} shows the transformation rules for the operations belonging to the point group $C_{3h}$. The constructed 2-band \textit{\textbf{k.p}} model for the K point is given by 
\begin{equation}
	H_\textrm{K}(\textbf{\textit{k}})=H_0(\textbf{\textit{k}})+k_x(3k_y^2-k_x^2)+\sigma_z[\Delta +\eta(k_x^2+k_y^2)+\zeta k_x(3k_y^2-k_x^2)]
\end{equation}

\subsection{M point}
\begin{table}[h]
	\begin{center}
		\caption{The transformations of ($\sigma_x$, $\sigma_y$, $\sigma_z$) and ($k_x$, $k_y$, $k_z$) with respect to the generators of the $C_{s}$ point group and time-reversal operator ($T$).}
		\begin{tabular}{p{3cm}p{4cm}p{4 cm}p{ 4 cm}}
			\hline
			\hline
			&  $M_{yz}=i\sigma_x$ & $M_{xy}=i\sigma_z$& $T=i\sigma_yK$           \\ \hline\hline
			
			$\;\;\;\,k_x$        &$\;-k_x$ & $\;\;\;\,k_x$& $-k_x$        \\ 
			$\;\;\;\,k_y$       & $\;\;\;\,k_y$ &$\;\;\;\,k_y$& $-k_y$      \\ 
			$\;\;\;\,k_z$       & $\;\;\;\,k_z$ &$\;-k_z$& $-k_z$       \\ 
			$\;\;\;\,\sigma_x$   & $\;\;\;\,\sigma_x$   &$\;-\sigma_x$& $-\sigma_x$   \\ 
			$\;\;\;\,\sigma_y$  & $\;-\sigma_y$  &$\;-\sigma_y$ & $-\sigma_y$   \\ 
			$\;\;\;\,\sigma_z$  & $\;-\sigma_z$   &$\;\;\;\,\sigma_z$& $-\sigma_z$  \\ \hline\hline
			& $k^m_ik_x\sigma_y$, $k^m_ik_x\sigma_z$,    & $k^m_ik_x\sigma_z$, $k^m_ik_y\sigma_z$,& $k_i\sigma_j $ \\
			Invariants              & $k^m_ik_y\sigma_x$, $k^m_ik_z\sigma_x$   &$k^m_ik_z\sigma_x$, $k^m_ik_z\sigma_y$ & ($i,j=x,y,z$) \\ 
			& ($i=x,y,z$; $m=0,2$)  & ($i=x,y,z$; $m=0,2$)& \\ \hline \hline
		\end{tabular}
		\label{tran3}
	\end{center}
\end{table}
The little point group for the M point is the $C_s$ containing identity operation ($E$), mirror operations perpendicular to principal axis ($\sigma_h$) and parallel to principal axis ($\sigma_v$). Employing the similar methodology as for $\Gamma$ and K points, the 2-band \textit{\textbf{k.p}} model is given by (see Table~\ref{tran3} for transformations) 
\begin{equation}
	H_\textrm{M}(\textbf{\textit{k}})=H_0(\textbf{\textit{k}}) + \alpha k_x \sigma_z
	\label{e3}
\end{equation}
\newpage
\section{Electric field effect on electronic structure}
\begin{figure}[h]
	\includegraphics[width=16cm]{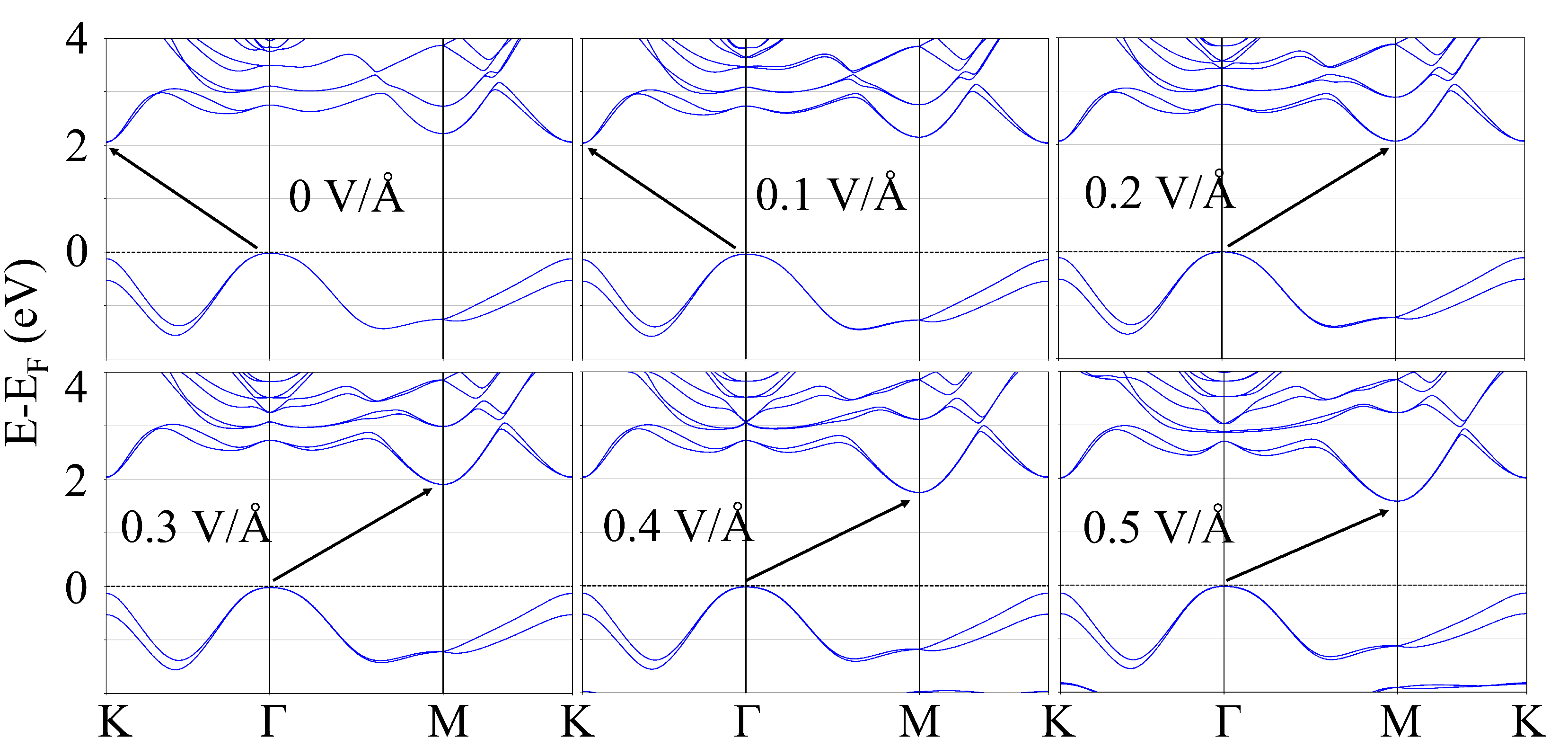}
	\caption{The evolution of PBE band structure of $\alpha_2$-WSi$_2$N$_4$ under the application of external electric field (EEF) in the range of 0 to 0.5 V/\AA. In the range 0 to 0.1 V/\AA, band gap is nearly independent of EEF having CBm at K point. However, transition of CBm from K to M point take place in the range of 0.1 to 0.2 V/\AA. Further, band gap shows linear and sharp decrement till EEF of 0.5 V\textrm/{\AA} (see Fig. 5 of main text). }
\end{figure}

\bibliography{siref}

\begin{figure}[h!]
\includegraphics[width=0.8\columnwidth,clip]{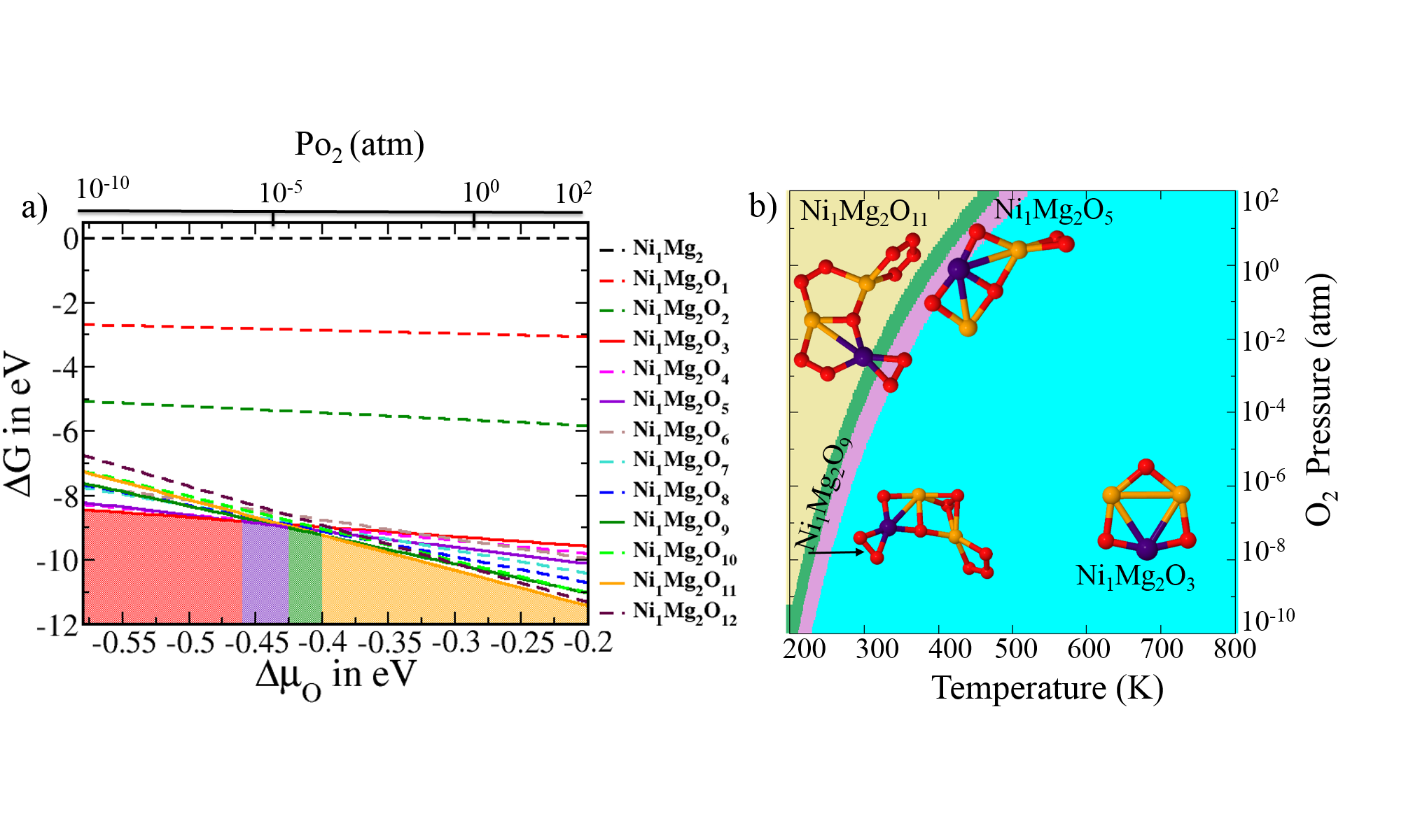}
\caption{In a), $\Delta G$ as a function of chemical potential of oxygen. Total energy of geometries are calculated using PBE+vdW. Colored areas are for the guideline to identify which phase is the most stable phase for different range of pressure. In b), 3D phase diagram for Ni$_1$Mg$_2$O$_x$ clusters}
\label{fig:SIpic1}
\end{figure}

\title{Small ternary $TM_x$Mg$_y$O$_z$ clusters ($TM$ = Cr, Ni, Fe, Co; $x+y \leq 3$) at realistic conditions: Unraveling stability and electronic structure from first-principles}
\author{Shikha Saini, Debalaya Sarker, Pooja Basera and Saswata Bhattacharya\footnote{saswata@physics.iitd.ac.in}} 
\affiliation{Ab initio Research Group, Department of Physics, Indian Institute of Technology Delhi, Hauz Khas 110016, New Delhi, India}
\date{\today}
\pacs{}
\keywords{CH$_3$NH$_3$PbI$_3$, CH$_3$NH$_3$SnI$_3$, defects, DFT, vacancy, carrier concentration, free energy}
\maketitle
\begin{center}
{\Large \bf Supplemental Material}\\ 
\end{center}
\vspace*{12pt}
\begin{enumerate}[\bf I.]

\item Effect of functionals on phase diagrams.

\end{enumerate}
\noindent {\bf \Large I. Effect of functionals on phase diagrams\\}
In the FIG. \ref{fig:SIpic1}a) below, we have plotted the free energy of formation $\Delta G$ of globally optimized clusters (Ni$_1$Mg$_2$O$_x$) with varying oxygen chemical potential ($\Delta {{\mu }_{{\rm O}}}$) at a finite temperature (T=300K). In order to evaluate $\Delta G$, we have calculated free energy of the clusters Ni$_1$Mg$_2$O$_x$ (for x=1,2,3...) that is approximated by their $E\textsuperscript{DFT}$ (DFT total energy) and $F\textsuperscript{vibrational}$ (see Eq. 1 and 2). Here, the electronic energies are calculated by using PBE+vdW functional. 
\begin{equation}
	\begin{split}
		\Delta G_f\left(T,p_{O_2}\right) = F_{Me_xMg_yO_z}(T) - F_{Me_xMg_y}(T) - z\Delta\mu_O\left(T,p_{O_2}\right)
		\label{eq2}
	\end{split}
\end{equation}
where,
\begin{equation}
	\begin{split}
		F(T) =  F\textsuperscript{vibrational}(T) + E \textsuperscript{DFT}
		\label{eq3}
	\end{split}
\end{equation}
The chemical potential of oxygen is calculated by using Eq. 3. Where, $\nu_{OO}$ is the O-O stretching frequency and $\frac{h\nu_{OO}}{2}$ is the zero point energy of the oxygen molecule.
\begin{equation}
	\begin{split}
		\Delta \mu_O \left(T,p_{O_2}\right) & =\frac{1}{2}\left(\mu_{O_2} \left(T,p_{O_2}\right) - E^\textsuperscript{DFT}_{O_2} - \dfrac{h\nu_{OO}}{2}\right)
		\label{eq4}
	\end{split}
\end{equation}
Where,
\begin{equation}
	\begin{split}
		\mu_{O_2}\left(T,p_{O_2}\right) & = -k_BT ln \left[\left(\frac{2\pi m}{h^2}\right)^\frac{3}{2} (k_BT)^\frac{5}{2}\right] + k_BT ln p_{O_2} - k_BT ln \left(\frac{8 \pi^2 I_Ak_BT}{h^2} \right)\\& + \frac{h\nu_{OO}}{2} + k_BT ln\left[ 1 - \exp \left(-\frac{h\nu_{OO}}{k_BT}\right)\right]+ E \textsuperscript{DFT} -k_BT ln \mathcal{M} 
		+k_BT ln \sigma 
		\label{eq5}
	\end{split}
\end{equation}

\begin{figure}[h!]
	\includegraphics[width=0.8\columnwidth,clip]{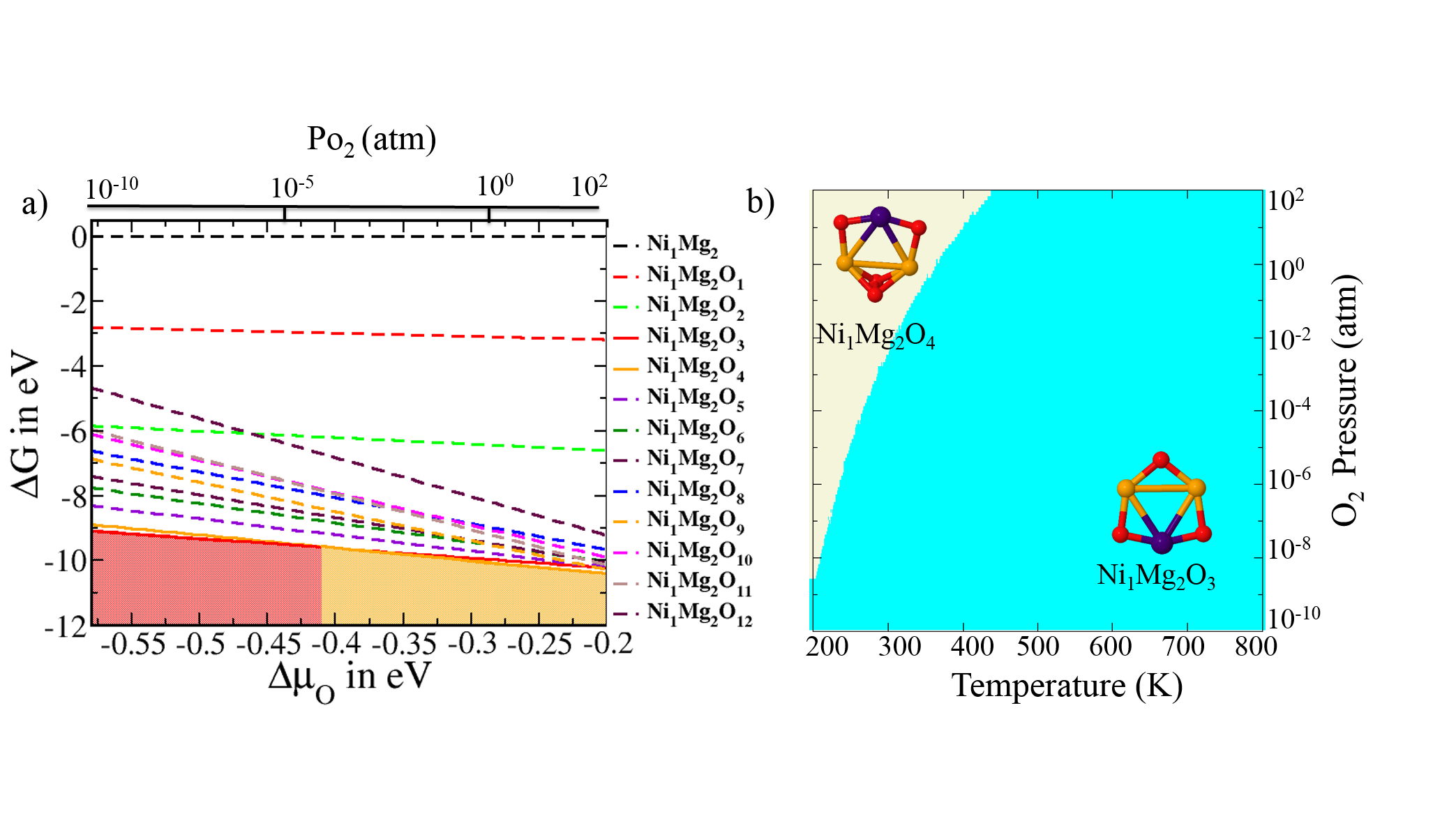}
	\caption{2D (a) and 3D (b) phase diagram for same set of clusters. Total energy of geometries are calculated using HSE06+vdW.}
	\label{fig:SIpic2}
\end{figure}

 In the above equation, $m$ is the mass, $I_A$ is the moment of inertia, $\mathcal{M}$ is the spin multiplicity, and $\sigma$ is the symmetry number for oxygen molecule. Pressure dependency in the free energy of formation $\Delta G$ comes in $\Delta\mu_O$ expression only. Therefore, the pressure axes are calculated according to the relation between $\mu_{O_2}$, with ${{p}_{{{{\rm O}}_{2}}}}$ as shown in Eq. 3 and 4. This variation of pressure in chemical potential ($\Delta {{\mu }_{{\rm O}}}$) is shown on top x-axis in the FIG. \ref{fig:SIpic1}a). The structures and compositions, which exhibit the lowest free energy of formation $\Delta G$ in the experimentally accessible (T, ${{p}_{{{{\rm O}}_{2}}}}$) region, is the most stable one under that (T, ${{p}_{{{{\rm O}}_{2}}}}$) region. In lower range of pressure Ni$_1$Mg$_2$O$_3$ is the most stable one, as we increase the pressure Ni$_1$Mg$_2$O$_5$, Ni$_1$Mg$_2$O$_9$ and, finally Ni$_1$Mg$_2$O$_{11}$ compositions are more preferable as shown in FIG. \ref{fig:SIpic1}a). Lowest lying lines are corresponding to the most stable phases as shown in FIG. \ref{fig:SIpic1}a). In FIG. \ref{fig:SIpic1}b), the dependence of temperature and pressure are combined in a 3D phase diagram. In 3D phase diagram, we could able only to concentrate on the stable phase with varying temperature and pressure simultaneously within single figure. In FIG. \ref{fig:SIpic1}b), there is a color corresponding to a particular composition, that indicates which phase is the most stable phase at which T and ${{p}_{{{{\rm O}}_{2}}}}$. In FIG. 1b) at high temperature and all range of pressure Ni$_1$Mg$_2$O$_3$ is the most stable phase, and at lower T and ${{p}_{{{{\rm O}}_{2}}}}$  Ni$_1$Mg$_2$O$_5$, Ni$_1$Mg$_2$O$_9$ are the most favorable. Further, as we increase the pressure at lower range od temperature Ni$_1$Mg$_2$O$_{11}$ phase is more preferable.
Same as in FIG. \ref{fig:SIpic1}, we have plotted the 2D and 3D phase diagrams by using more advanced hybrid functional (HSE06+vdW). Here, we have calculated the electronic energies of Ni$_1$Mg$_2$O$_x$ clusters by using HSE06+vdW functional. From the comparison of FIG. \ref{fig:SIpic1} and FIG. \ref{fig:SIpic2}, it can be clearly seen that there are huge difference in the phase diagrams with different functionals. In FIG. \ref{fig:SIpic2}a), at lower pressure range Ni$_1$Mg$_2$O$_3$ is the most stable phase at 300K. As we increase the pressure Ni$_1$Mg$_2$O$_4$ is the favorable phase. From FIG. \ref{fig:SIpic2}b) we can see the most stable phases at all range of temperature with varying the pressure. At lower range of pressure and all temperature Ni$_1$Mg$_2$O$_3$ is most stable one and as we increase the pressure at lower range of temperature Ni$_1$Mg$_2$O$_4$ is the favorable phase.  
\bibliography{Bibliography}{}